\documentclass[journal=jctcce,manuscript=article]{achemso}

\usepackage[version=3]{mhchem} %

\usepackage[utf8]{inputenc}
\usepackage[dvipsnames]{xcolor}
\usepackage{tikz}
\usepackage{amsmath}
\usepackage{graphicx}
\usepackage{subcaption}
\usepackage{braket}
\usepackage{tabularx}

\usetikzlibrary{shapes.geometric, arrows}

\newcounter{algorithm}
\renewcommand{\thealgorithm}{\arabic{algorithm}}
\newenvironment{pseudocode}[1]{%
  \begin{figure}[htbp]%
  \noindent\begin{minipage}[t]{\textwidth}%
  \refstepcounter{algorithm}%
  \rule{\textwidth}{0.8pt}\\[2pt]
  \textbf{Algorithm \thealgorithm}\quad #1\\[2pt]
  \rule{\textwidth}{0.4pt}%
  \small\setstretch{1.0}%
}{%
  \par\noindent\rule{\textwidth}{0.8pt}%
  \end{minipage}%
  \end{figure}%
}

\title{Surface Hopping Dynamics on Adiabatic States Propagated with Real-Time TDDFT}

\author{Ruiyi Zhou}
\affiliation{Institute of Molecular Physical Science, ETH Zürich, 8093 Zürich, Switzerland}
\email{ruiyi.zhou@phys.chem.ethz.ch}
\author{Jeremy O. Richardson}
\affiliation{Institute of Molecular Physical Science, ETH Zürich, 8093 Zürich, Switzerland}
\email{jeremy.richardson@phys.chem.ethz.ch}

\begin{document}
\begin{abstract}
Real-time time-dependent density functional theory (RT-TDDFT) is widely used to simulate ultrafast electron dynamics, but it is usually coupled to the nuclei through the mean-field Ehrenfest approximation, which cannot describe wavepacket branching and violates detailed balance. Surface hopping avoids these problems, but has not previously been combined with RT-TDDFT because real-time propagation does not directly provide the necessary adiabatic states. In this work we introduce an energy-scaled time-dependent Kohn--Sham (TDKS) propagation scheme, motivated by the quantum adiabatic theorem. Scaling widens the separation between electronic and nuclear timescales, so that each propagated state follows the adiabatic surface selected by its initial condition: either the ground state or a non-Aufbau self-consistent-field ($\Delta$SCF) solution. The propagation thereby avoids the convergence failures of static SCF solvers in regions of strong nonadiabatic coupling. Because the independently propagated states are complex-valued and not mutually orthogonal, we reformulate the mapping approach to surface hopping (MASH) in a complex adiabatic basis and introduce a dynamical transition orbital (DTO) gauge that restores a well-defined particle--hole representation for evaluating time-derivative couplings and nonadiabatic coupling vectors. Applied to the photoisomerization of \textit{trans}-diimide, the scheme converges systematically as the scaling factor is increased and yields an excited-state lifetime in good agreement with previous nonadiabatic simulations based on static SCF.\@ These results establish real-time propagation as an efficient and numerically stable route to the adiabatic states and couplings required for first-principles surface hopping.
\end{abstract}

\maketitle

\begin{tocentry}
\includegraphics[width=\textwidth]{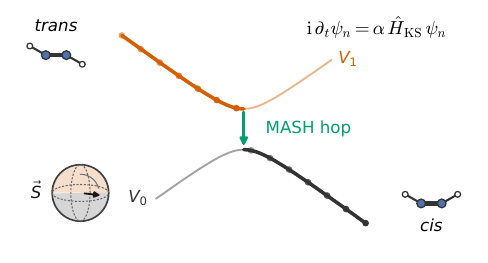}
\end{tocentry}

\section{Introduction}

Simulating photoinduced chemical reactions requires going beyond the Born--Oppenheimer approximation because photoexcited molecules readily access regions of strong nonadiabatic coupling such as the vicinity of conical intersections.\cite{NonadiabaticBook} The simplest nonadiabatic dynamics scheme employs the Ehrenfest approximation, which propagates the nuclei on a mean-field average of the potential-energy surfaces.\cite{li2005ab} However, its mean-field character prevents nuclear trajectories from branching onto distinct product surfaces\cite{Tully2012perspective} and can lead to severe violations of detailed balance.\cite{Parandekar2006Ehrenfest} Surface-hopping methods, such as fewest-switches surface hopping (FSSH)\cite{tully1990molecular} and the mapping approach to surface hopping (MASH),\cite{MASH,MASHreview} instead propagate each trajectory on a single active adiabatic surface and define a rule for transitioning between states. In this way, an ensemble of surface-hopping trajectories can branch onto different surfaces after passing through a nonadiabatic coupling region, which typically leads to more reliable results than Ehrenfest dynamics.

Most surface-hopping methods require the adiabatic electronic states $\Phi_N$ and the energies $E_N$, forces and nonadiabatic couplings derived from them.\cite{Mai2018SHARC,crespo2018recent}
These states are formally defined as the eigenstates of the electronic Hamiltonian, %
\begin{equation}\label{eq:adiabatic_basis_intro}
    \hat{H}_{\mathrm{el}}\Phi_N = E_N\Phi_N
\end{equation}
For model Hamiltonians, the states are simply calculated by matrix diagonalization, whereas obtaining them on the fly from first principles is considerably more demanding. Multireference wavefunction methods such as complete-active-space second-order perturbation theory (CASPT2)\cite{andersson1992second} and multireference configuration interaction (MRCI)\cite{szalay2012multiconfiguration} can describe excited states and conical intersections accurately, but their computational cost restricts the accessible system size. Density functional theory (DFT),\cite{hohenberg1964inhomogeneous,parr1995density} with the Kohn--Sham (KS) representation of the density by an auxiliary noninteracting determinant,\cite{kohn1965self} provides a more affordable starting point for the ground state. The challenge is to obtain the energies, forces and couplings of excited states reliably as the nuclei move through strongly coupled regions.

DFT offers several ways to treat excited states within a static picture, in which calculations are performed independently at each time step. Linear-response time-dependent density functional theory (LR-TDDFT),\cite{casida1995time,UllrichBook} which supplies a manifold of excitation energies and response states, is widely used with surface hopping.\cite{tapavicza2007trajectory} Its cost is, however, higher than that of ground-state KS-DFT, and with commonly used adiabatic exchange--correlation kernels, it does not describe intersections involving the ground state correctly.\cite{levine2006conical}

The $\Delta$SCF method instead approximates an excited state by a separate self-consistent-field (SCF) calculation with a non-Aufbau occupation,\cite{yang2024foundation} at a cost comparable to a ground-state calculation, and is often more accurate than LR-TDDFT for charge-transfer, doubly excited states and conical intersections.\cite{hait2020excited, schmerwitz2022variational,vigneshwaran2026variational} Although $\Delta$SCF has been combined with surface hopping,\cite{pradhan2018non,hanasaki2025exploring,vandaele2022delta} variational collapse\cite{allen1987examination} and switching between competing SCF solutions remain practical difficulties. The most widely used strategy is the maximum overlap method (MOM),\cite{gilbert2008self} which at each SCF iteration occupies the orbitals that overlap most with those of the previous iteration. Other optimizers target individual excited states directly.\cite{hait2020excited,carter2020state,kumar2022robust,levi2020variational} Along a trajectory, however, the overlap criterion must be reapplied at every geometry, and near degeneracies, where several orbitals overlap comparably with the reference, it can incorrectly switch to a different solution. Related state-specific approaches include restricted open-shell Kohn--Sham (ROKS),\cite{frank1998molecular,hait2016prediction} $\sigma$-SCF,\cite{ye2017sigma} and constrained DFT.\cite{ziegler2009relation,evangelista2013orthogonality,ramos2018low} %

In contrast to these static, state-specific approaches, real-time TDDFT (RT-TDDFT) propagates the KS orbitals directly in the time domain,\cite{runge1984density,marques2012fundamentals,yabana1996time} giving access to strongly nonequilibrium processes\cite{li2020real,xu2024real,ullrich2025snapshot,zhou2021first} and to linear and nonlinear optical response.\cite{provorse2016electron}
Real-time propagation can be more efficient and more stable than the static SCF approaches discussed above.\cite{alonso2008efficient,andrade2009modified}
The propagated density does not directly supply separate adiabatic-state energies, forces and couplings, however, so coupled electron--nuclear applications have largely relied on Ehrenfest dynamics.\cite{castro2004excited,li2005ab,meng2008real,wang2011time,Alonso2012TDDFT,ojanpera2012nonadiabatic,kolesov2016real,bonafe2020real,shepard2021simulating,xu2023first} Extensions beyond this level differ mainly in what takes the place of the adiabatic basis: natural-orbital branching,\cite{wang2020natural} density-matrix ensembles with decoherence corrections,\cite{kang2019nonadiabatic} and approaches based on exact factorization.\cite{abedi2010exact,ha2022independent,han2023real} Another widely used approach describes transitions between instantaneous eigenstates of the time-dependent Kohn--Sham (TDKS) Hamiltonian,\cite{craig2005trajectory,fischer2011regarding} in an independent-electron picture also used for surface hopping at metal surfaces.\cite{shenvi2009nonadiabatic} However, these single-particle states are not the same as the many-electron adiabatic states of Eq.~\ref{eq:adiabatic_basis_intro}. Implementations such as PYXAID,\cite{akimov2013pyxaid} Hefei-NAMD,\cite{zheng2019ab} and GTSH\cite{liu2023nonadiabatic} have enabled simulations of exciton and carrier relaxation in condensed-phase systems.\cite{wang2016recent} For simplicity, many such calculations use the classical-path approximation, in which the nuclei follow a ground-state trajectory irrespective of the active electronic state.\cite{akimov2013pyxaid} However, photochemistry involving large nuclear rearrangement, such as photoisomerization, requires feedback from nuclear forces resulting from the many-electron adiabatic states.

The concept of propagating electronic states to follow adiabatic surfaces has precedents in Car--Parrinello molecular dynamics (CPMD)\cite{car1985unified,MarxHutterBook,hutter2012car,kuhne2014second} and other extended-Lagrangian molecular dynamics,\cite{niklasson2008extended} originally developed for ground-state Born--Oppenheimer dynamics. In CPMD, a fictitious mass controls the separation between electronic and nuclear timescales required for adiabatic following.\cite{pastore1991theory} %
Extensions have been used for propagation on excited-state surfaces.\cite{frank1998molecular,bjorgaard2018extended,fajen2025shadow,tawfik2025hole} For instance, the nonadiabatic CPMD scheme of Doltsinis and Marx\cite{doltsinis2002nonadiabatic} propagates a ROKS excited state alongside a restricted Kohn--Sham ground state and evaluates their coupling from the finite-difference overlap expression of Hammes-Schiffer and Tully.\cite{hammes1994proton} However, this approach assumes orthonormal states, whereas independently generated KS states need not be mutually orthogonal. For open-shell states, mixing of equally occupied active spatial orbitals can also change the constructed many-electron wavefunction without changing the density, as discussed in studies of ROKS.\cite{kowalczyk2013excitation,buchel2024simulation} A propagation-based surface-hopping scheme therefore needs both reliable following of the selected SCF branches and a consistent wavefunction representation for evaluating their overlaps and couplings.

Here we combine energy-scaled TDKS propagation with a complex-state formulation of MASH\@. Motivated by the quantum adiabatic theorem, we multiply the electronic Hamiltonian by a factor $\alpha>1$ and propagate each state independently. For a prescribed nuclear path, this scaling is equivalent to traversing the same path at $1/\alpha$ of the original speed, so that each propagated state follows its adiabatic state more closely (Sec.~\ref{sec:following_theory}).
Note that a similar electronic equation of motion underlies the modified Ehrenfest dynamics of Rubio and coworkers, which uses the opposite scaling regime to permit larger time steps in ground-state simulations.\cite{alonso2008efficient,andrade2009modified}
In order to recover the nonadiabatic dynamics that we wish to describe, we use MASH
with unscaled energies and forces.
MASH is a variant of surface hopping that uses a deterministic algorithm for selecting the active surface
and shows a number of benefits over the original FSSH approach.\cite{MASH,MASHreview}
Nonetheless, the implementation of MASH is very similar to that of other surface-hopping approaches: they require energies, forces and time-derivative couplings at each time step, as well as nonadiabatic coupling vectors at hops.
Since real-time propagation produces complex states, we recast the MASH equations in a complex adiabatic basis. A dynamical transition orbital (DTO)\cite{zhou2021dynamical} gauge fixes the particle--hole representation used to construct nearly orthogonal many-electron wavefunctions. Their overlaps then provide the required time-derivative couplings and nonadiabatic coupling vectors.

We demonstrate the scheme for the photoisomerization of \textit{trans}-diimide, testing state following, wavefunction overlaps and nonadiabatic couplings before performing full \mbox{RT-TDDFT/MASH} dynamics. Increasing $\alpha$ systematically improves state following through regions where static SCF optimization is unreliable, and the ensemble populations and structural observables converge to results in good agreement with earlier surface-hopping simulations based on static $\Delta$SCF calculations.\cite{malis2020trajectory} In this proof-of-principle study, the main demonstrated advantages are robustness and numerical stability. For larger systems, where static excited-state SCF calculations become increasingly difficult to converge, the approach is also expected to be more efficient than standard static approaches (as observed in other RT-TDDFT studies of large systems),\cite{alonso2008efficient,andrade2009modified} although a systematic cost comparison is beyond the scope of this work.

We describe the electronic propagation in Sec.~\ref{sec:rttddft}, the coupled nonadiabatic dynamics in Sec.~\ref{sec:sh}, test the method in Sec.~\ref{sec:results}, and conclude in Sec.~\ref{sec:conclusions}.

\section{RT-TDDFT for Adiabatic States}\label{sec:rttddft}

\emph{Notation.} We first fix the conventions used throughout. Matrices are set in upright boldface (e.g., $\mathbf{S}$, $\mathbf{H}$, $\mathbf{C}$) and vectors in italic boldface (e.g., $\boldsymbol{R}$, $\boldsymbol{P}$, $\boldsymbol{d}$). The letter $S$ is conventionally used both for the atomic-orbital overlap matrix of the LCAO equations of Sec.~\ref{sec:lcao} and for the spin (Bloch) vector of the MASH mapping variables of Sec.~\ref{sec:mash_complex}. We retain both usages but write the latter with an arrow, $\vec{S}$, with Cartesian components $S_x$, $S_y$, $S_z$, so that the two are distinguished at a glance. For the same reason, the orbital-overlap matrices used to evaluate many-electron overlaps in Sec.~\ref{sec:nac} are denoted $\mathbf{O}$ rather than the more common $\mathbf{S}$. Atomic units are used unless stated otherwise.

\subsection{Adiabatic States from the Variational Principle}\label{sec:variational}

Within the framework of Kohn--Sham density functional theory (KS-DFT),\cite{kohn1965self,parr1995density} the adiabatic ground state for a fixed nuclear configuration $\boldsymbol{R}$ is obtained by minimizing the total electronic energy functional $E_{\mathrm{KS}}[\rho]$, which is the sum of the noninteracting kinetic energy, the external-potential energy, the classical Hartree energy, and the exchange--correlation energy,\cite{parr1989density} %
\begin{equation}\label{eq:eks}
    E_{\mathrm{KS}}[\rho;\boldsymbol{R}] = T_{\mathrm{s}}[\{\psi_n\}] + \int v_{\mathrm{ext}}(\boldsymbol{r};\boldsymbol{R})\rho(\boldsymbol{r})\, \mathrm{d}^3\boldsymbol{r}  + \frac{1}{2} \iint \frac{\rho(\boldsymbol{r})\rho(\boldsymbol{r}')}{|\boldsymbol{r} - \boldsymbol{r}'|}\, \mathrm{d}^3\boldsymbol{r}\, \mathrm{d}^3\boldsymbol{r}' + E_{\mathrm{xc}}[\rho]
\end{equation}
Here, the density $\rho(\boldsymbol{r})$ is a function of a single-electron coordinate $\boldsymbol{r}$ and is constructed from the occupied KS orbitals:
\begin{equation}\label{eq:electron_density}
    \rho(\boldsymbol{r}) = \sum_{n} f_n |\psi_n(\boldsymbol{r})|^2
\end{equation}
where $f_n$ is the occupation number of orbital $n$. Orthonormality of the orbitals is imposed during the minimization through Lagrange multipliers $\varepsilon_{nm}$,
\begin{equation}
    \mathcal{L} [\{\psi_n\};\boldsymbol{R}]= E_{\mathrm{KS}}[\rho;\boldsymbol{R}] - \sum_{n,m} \varepsilon_{nm}(\langle \psi_m | \psi_n \rangle - \delta_{mn})
\end{equation}
Variation with respect to the orbitals gives the KS self-consistent-field (SCF) equations
\begin{equation}\label{eq:ks}
    \hat{H}_{\mathrm{KS}}[\rho] \psi_n = \varepsilon_n \psi_n
\end{equation}
where $\hat{H}_{\mathrm{KS}}[\rho] = -\tfrac{1}{2}\nabla^2 + \hat{v}_{\mathrm{ext}}(\boldsymbol{R}) + \hat{v}_{\mathrm{H}}[\rho] + \hat{v}_{\mathrm{xc}}[\rho]$ is the KS Hamiltonian.
Because $\hat{H}_{\mathrm{KS}}$ depends on $\rho$ and hence on the KS orbitals themselves, Eq.~\ref{eq:ks} is a nonlinear eigenvalue problem that must be solved iteratively. The resulting potential-energy surface on which the nuclei move is $V(\boldsymbol{R})=E_{\mathrm{KS}}[\rho;\boldsymbol{R}]+E_{\mathrm{nuc}}(\boldsymbol{R})$, where $E_{\mathrm{nuc}}$ is the nuclear repulsion energy.

In practical implementations, a range of numerical techniques is used to accelerate the convergence of the SCF iteration of Eq.~\ref{eq:ks}. Direct inversion in the iterative subspace (DIIS)\cite{pulay1980convergence} extrapolates the KS Hamiltonian from previous iterations, whereas orbital-optimization methods parametrize the orbitals by a unitary rotation $\mathrm{e}^{\boldsymbol{\kappa}}$ and minimize the energy functional directly with respect to the anti-Hermitian matrix $\boldsymbol{\kappa}$ using its gradient and an exact or approximate Hessian.\cite{head1988optimization,van2002geometric} Because every step lowers the energy, these methods converge reliably to a minimum, which makes them robust for ground states. $\Delta$SCF excited states, however, are typically saddle points of the energy and can collapse to lower solutions unless the optimization is modified.\cite{levi2020variational,hait2020excited} Throughout this work, we generate static SCF reference data with both standard DIIS and the second-order co-iterative augmented-Hessian (CIAH) orbital optimizer,\cite{sun2016co} as implemented in PySCF~\cite{sun2018pyscf,sun2020recent}.

Neither strategy completely removes the difficulties that static SCF poses in a molecular-dynamics setting. Converging a full SCF calculation at every time step is expensive, and convergence itself becomes unreliable in the strong-coupling regions that govern nonadiabatic transitions. In particular, near avoided crossings, where orbital energies are close to degenerate, the SCF iteration may oscillate between competing configurations or fail outright, producing discontinuous energies and forces.

\subsection{The Ground and $\Delta$SCF Excited States}\label{sec:target_states}

In this work, we restrict ourselves to two states: the ground state ($N=0$) and the lowest singly excited singlet ($N=1$). We define them here in the static picture, that is, as stationary SCF solutions converged at a fixed nuclear geometry, before turning to their dynamical propagation.  Both are described in a spin-restricted formalism throughout. Applications involving bond breaking or pronounced open-shell character will require an extension to unrestricted Kohn--Sham theory, which we leave to future work.

\begin{itemize}
    \item \textbf{Ground state ($N=0$).} The standard Aufbau solution, in which the lowest orbitals are doubly occupied ($f_n=2$) and the remaining orbitals are empty ($f_n=0$), as depicted in Fig.~\ref{fig:ground_state}. The many-electron Kohn--Sham (KS) wavefunction is the single Slater determinant
    \begin{equation}\label{eq:SD_gs}
        \Phi_0 = |\psi_1 \bar{\psi}_1 \dots \psi_{\mathrm{HOMO}}\bar{\psi}_\mathrm{HOMO}\rangle
    \end{equation}

    \item \textbf{Excited state ($N=1$).} A non-Aufbau configuration in which one electron is promoted from the HOMO to the LUMO (Fig.~\ref{fig:excited_state}). Formally, the spin-restricted open-shell singlet is the two-determinant combination
    \begin{equation}\label{eq:SD_es}
        \Phi_1 = \frac{1}{\sqrt{2}} \left( |\psi_1 \bar{\psi}_1 \dots \psi_{\mathrm{HOMO}} \bar{\psi}_{\mathrm{LUMO}}\rangle - |\psi_1 \bar{\psi}_1 \dots \bar{\psi}_{\mathrm{HOMO}} \psi_{\mathrm{LUMO}}\rangle \right)
    \end{equation}
    However, we work throughout with a spin-restricted, occupation-constrained $\Delta$SCF description,\cite{ziegler1977calculation} in which a single set of restricted spatial orbitals is used, with occupation numbers $f_n=1$ for the two singly occupied orbitals (HOMO and LUMO) and $f_n=2$ for the remaining occupied orbitals. The energies and the corresponding nuclear forces are therefore all obtained from the spin-restricted KS energy functional evaluated on the density of Eq.~\ref{eq:electron_density} built with these occupation numbers, that is, with a spin-unpolarized density. This energy differs both from that of the spin-resolved mixed determinant and from the spin-purified singlet energy of ROKS,\cite{frank1998molecular,hait2016prediction} and we use it consistently in the static and dynamical calculations. Equation~\ref{eq:SD_es} is used only to define the spin-adapted many-electron wavefunction $\Phi_1$ entering the state overlaps of Sec.~\ref{sec:nac}, where it determines the form of Eq.~\ref{eq:U_det}.
\end{itemize}

The adiabatic states are calculated differently in the static and dynamical settings, and the distinction matters for the comparisons made later. For the static reference calculations, the maximum-overlap method (MOM)\cite{gilbert2008self} is applied at \emph{every} geometry along the trajectory. In the dynamical scheme, by contrast, MOM is used only once, to converge the state at the initial time, and thereafter the excited-state character is maintained by the energy-scaled propagation itself, through the continuity of the time evolution. Dispensing with the orbital-selection criterion at every step is one of the practical advantages of the present approach, and, as we show later, it helps the propagated trajectory stay on the same branch even when the static solver jumps to a different root (Sec.~\ref{sec:following}).

\begin{figure}[htbp]
    \centering
    \begin{subfigure}[t]{0.45\textwidth}
        \centering
        \begin{tikzpicture}[level/.style={draw, thick}, electron/.style={fill, circle, inner sep=1.5pt}]
            \foreach \y/\label in {1.2/LUMO+1, 0/LUMO, -1.2/HOMO, -2.4/{HOMO$-$1}, -4/{1}} {
                \draw[level] (-1, \y) -- (1, \y);
                \node[right=1mm] at (1, \y) {\footnotesize $\psi_{\text{\label}}$};
            }
            \node at (0, -3.2) {\vdots};
            \foreach \y in {-1.2, -2.4, -4} {
                \node[electron] at (-0.3, \y) {};
                \node[electron] at (0.3, \y) {};
            }
        \end{tikzpicture}
        \subcaption{Ground state ($N=0$)}
        \label{fig:ground_state}
    \end{subfigure}
    \hfill
    \begin{subfigure}[t]{0.45\textwidth}
        \centering
        \begin{tikzpicture}[level/.style={draw, thick}, electron/.style={fill, circle, inner sep=1.5pt}]
            \foreach \y/\label in {1.2/LUMO+1, 0/LUMO, -1.2/HOMO, -2.4/{HOMO$-$1}, -4/{1}} {
                \draw[level] (-1, \y) -- (1, \y);
                \node[right=1mm] at (1, \y) {\footnotesize $\psi_{\text{\label}}$};
            }
            \node at (0, -3.2) {\vdots};
            \node[electron] at (0, 0) {};    %
            \node[electron] at (0, -1.2) {}; %
            \foreach \y in {-2.4, -4} {
                \node[electron] at (-0.3, \y) {};
                \node[electron] at (0.3, \y) {};
            }
        \end{tikzpicture}
        \subcaption{First excited state ($N=1$)}
        \label{fig:excited_state}
    \end{subfigure}
    \caption{Molecular-orbital occupation patterns of (a) the ground state ($N=0$) and (b) the first excited state ($N=1$).}
    \label{fig:mo_diagrams}
\end{figure}

\subsection{Adiabatic-State Following via Real-Time Propagation}\label{sec:following_theory}
The convergence problems near degeneracies stem from treating each geometry as an independent variational problem. We therefore replace repeated static optimizations by a dynamical propagation scheme, in which the target state is carried from one geometry to the next by the continuity of its own time evolution. This scheme rests on the quantum adiabatic theorem.

\subsubsection{The Quantum Adiabatic Theorem}\label{sec:adiabatic_theorem}

The quantum adiabatic theorem states that a system initialized in an eigenstate of a time-dependent Hamiltonian $\hat{H}(t)$ remains in the corresponding instantaneous eigenstate $\Phi_N(t)$ provided the Hamiltonian changes slowly on the characteristic electronic timescale.~\cite{born1928beweis,kato1950adiabatic,griffiths2018introduction} In other words, the timescale of the external perturbation, here driven by the nuclear motion, must be long compared with the electronic response time $\tau_{\text{el}}\sim1/\Delta E$, where $\Delta E$ is the relevant energy gap.  To be more quantitative, one can use Landau--Zener theory to provide estimates of whether a transition will be adiabatic given the velocity of the nuclei.\cite{Landau1932LZ,Zener1932LZ}

When this separation of timescales holds, RT-TDDFT coupled to Ehrenfest dynamics follows an approximately adiabatic trajectory, much as Born--Oppenheimer molecular dynamics does.\cite{miyamoto2015conservation} However, at high nuclear velocities or for small energy gaps, as near an avoided crossing, the adiabatic condition fails, and the electronic wavefunction develops into a superposition of adiabatic states. The forces resulting from such a state are a mean field over the adiabatic states and thus lead to the Ehrenfest description of nonadiabatic dynamics.

In order to go beyond the mean-field description, surface hopping requires pure adiabatic states with state-specific forces. Standard RT-TDDFT is therefore not directly suited to generating the underlying adiabatic surfaces. Within the time-dependent formalism, we therefore introduce an energy-scaling parameter $\alpha>1$ through the transformation $\hat{H}_{\mathrm{KS}} \mapsto \alpha \hat{H}_{\mathrm{KS}}$, which raises the frequency of the electronic phase evolution relative to the nuclear motion and so widens the separation of the two timescales and recovers the adiabatic condition.

\subsubsection{Energy-Scaled Equations of Motion}

In this section, we focus on the electronic dynamics and leave the formulation of the coupled electron--nuclear nonadiabatic dynamics to Sec.~\ref{sec:sh}. We therefore consider a prescribed nuclear trajectory $\boldsymbol{R}(t)$, for which we define the electronic action over the interval $[t_0,t_1]$,
\begin{equation}
    \mathcal{A} = \int_{t_0}^{t_1} \mathrm{d}t \left\{ \sum_{n} f_n \left\langle \psi_n(t) \left| \mathrm{i}\frac{\partial}{\partial t} \right| \psi_n(t) \right\rangle - \alpha E_{\mathrm{KS}}[\rho(t);\boldsymbol{R}(t)] \right\}
\end{equation}
where $E_{\mathrm{KS}}$ is the functional of Eq.~\ref{eq:eks} evaluated on the instantaneous density $\rho(t)$, built from the propagated TDKS orbitals with fixed occupation numbers. The initial orbitals $\{\psi_n(t_0)\}$ are taken to be a stationary converged solution of Eq.~\ref{eq:ks}. Making this action stationary yields the energy-scaled TDKS equations
\begin{equation} \label{eq:tdks_eom}
    \mathrm{i} \frac{\partial}{\partial t} |\psi_{n}(t)\rangle = \alpha \hat{H}_{\mathrm{KS}}(t)|\psi_{n}(t)\rangle
\end{equation}
with $\hat{H}_{\mathrm{KS}}(t)\equiv\hat{H}_{\mathrm{KS}}[\rho(t);\boldsymbol{R}(t)]$.

Setting $\alpha=1$ recovers the standard real-time TDKS equations. The introduction of the scaling parameter $\alpha$ can be interpreted as a change of the electronic clock. Defining the rescaled time $s=\alpha t$, Eq.~\ref{eq:tdks_eom} becomes
\begin{equation}\label{eq:tau_reparam}
    \mathrm{i}\frac{\partial}{\partial s}|\psi_n(s)\rangle = \hat{H}_{\mathrm{KS}}\!\left[\rho(s);\boldsymbol{R}(s/\alpha)\right]|\psi_n(s)\rangle
\end{equation}
which is the standard, unscaled TDKS equation propagated along the \emph{same} nuclear path traversed at $1/\alpha$ of the original speed. Scaling the electronic Hamiltonian by $\alpha$ is therefore equivalent to slowing the nuclei by the same factor. The equivalence is exact and survives the nonlinearity of $\hat{H}_{\mathrm{KS}}$ because $\rho$ is built from the instantaneous TDKS orbitals and is reparametrized along with them. Note that the scaling enters only the orbital equation of motion. All energies and nuclear forces, including the potential energies entering MASH, are evaluated from the unscaled KS energy functional on the propagated density, so the potential-energy surfaces themselves are unchanged.

Through Eq.~\ref{eq:tau_reparam}, the adiabatic theorem of Sec.~\ref{sec:adiabatic_theorem} suggests that in the $\alpha\to\infty$ limit, where the electrons experience an infinitely slow nuclear motion, each propagated state rigorously recovers the instantaneous adiabatic state selected by its initial condition, provided this state stays separated from the others by a finite gap. %
However, if the gap is zero, it is clear that no finite value of $\alpha$ can be found which obeys the adiabatic principle.
Such a case is called a trivial crossing and
there is nonetheless a simple remedy, due to the fact that the propagated states will retain their electronic character while exchanging their energy ordering.
Thus the correct ordering can be easily restored by
state tracking and surface switching.\cite{granucci2001direct,wang2014simple,mannouch2026ensuring} No trivial crossings occurred in the trajectories reported in this work.

We note that the electronic equation of motion, Eq.~\ref{eq:tdks_eom}, is mathematically equivalent to that of the modified Ehrenfest dynamics of Rubio and coworkers,\cite{alonso2008efficient,andrade2009modified} $\mathrm{i}\mu\,\partial_t\psi_n=\hat{H}_{\mathrm{KS}}\psi_n$, with $\alpha=1/\mu$. The two methods, however, use this parameter for different purposes. Modified Ehrenfest dynamics was designed as an efficient alternative to ground-state Born--Oppenheimer molecular dynamics and takes $\mu\gg1$, that is, $\alpha\ll1$. The electronic excitation energies are then scaled down by $1/\mu$ and the time step can be increased by a factor of $\mu$, at the price of weaker adiabaticity, so the admissible $\mu$ is limited by the electronic gap. Those studies also discussed the adiabatic limit at small $\mu$. Here we work at $\alpha>1$ and accept a correspondingly smaller time step in exchange for adiabatic following, with $\alpha$ chosen large enough to obtain numerical convergence. As demonstrated in Sec.~\ref{sec:following}, this lets us follow $\Delta$SCF excited states as well as the ground state through regions where static SCF fails, thereby providing the adiabatic states and couplings that surface hopping requires. In practice, Eq.~\ref{eq:tdks_eom} differs from the standard RT-TDDFT equation only by the scalar factor $\alpha$, so the energy scaling itself requires only a minor modification of the orbital propagation in an existing RT-TDDFT Ehrenfest code.

\subsection{Numerical Implementation in an Atom-Centered Orbital Basis}\label{sec:lcao}

We represent the time-dependent KS orbitals in a basis of atomic orbitals (AOs) using the linear-combination-of-atomic-orbitals (LCAO) expansion
\begin{equation}
    \psi_n(\boldsymbol{r}, t) = \sum_{\mu} c_{\mu n}(t) \phi_{\mu}(\boldsymbol{r} - \boldsymbol{R}_A(t))
\end{equation}
such that each basis function $\phi_{\mu}$ is centered on its corresponding nucleus $A$ at its instantaneous position $\boldsymbol{R}_A(t)$. The basis functions thus acquire an implicit time dependence through the nuclear motion. In this way, the LCAO expansion employs a moving diabatic basis of local functions attached to their nuclei.

Substituting the expansion into Eq.~\ref{eq:tdks_eom} and projecting onto the atomic basis gives the matrix equation of motion for the coefficients. Because the basis functions are themselves time dependent, the chain rule applied to $\partial_t\psi_n$ produces two terms, $\sum_\mu\left(\dot{c}_{\mu n}\phi_\mu + c_{\mu n}\dot{\phi}_\mu\right)$, the second of which has no counterpart in a fixed basis. Projecting onto $\phi_\mu$ we obtain

\begin{equation} \label{eq:EOM_wf}
    \sum_{\nu} \mathbf{S}_{\mu \nu} \dot{c}_{\nu n} = - \sum_{\nu} \left( \mathrm{i} \alpha \mathbf{H}_{\mu \nu} + \mathbf{D}_{\mu \nu} \right) c_{\nu n}
\end{equation}
where the matrix elements are
\begin{subequations}
\begin{align}
    \mathbf{S}_{\mu \nu} &= \braket{\phi_\mu(t) | \phi_\nu(t)} \\
    \mathbf{H}_{\mu \nu} &= \braket{\phi_\mu(t) | \hat{H}_{\mathrm{KS}}(t) | \phi_\nu(t)} \\
    \mathbf{D}_{\mu \nu} &= \braket{\phi_\mu(t) | \dot{\phi}_\nu(t)}
     = \sum_{A} \dot{\boldsymbol{R}}_A(t) \cdot \braket{\phi_\mu(\boldsymbol{R}(t)) | \nabla_A \phi_\nu(\boldsymbol{R}(t))} \label{eq:tdks_nuclei_coupling}
\end{align}
\end{subequations}
Here $\mathbf{S}$ and $\mathbf{H}$ are the AO overlap and KS Hamiltonian matrices, and $\mathbf{D}$ is the time-derivative coupling induced by the motion of the basis functions, expressed in Eq.~\ref{eq:tdks_nuclei_coupling} in terms of the nuclear velocities $\dot{\boldsymbol{R}}_A$ by the chain rule, with $\nabla_A$ the gradient with respect to the position coordinate $\boldsymbol{R}_A$ of nucleus $A$. Because the basis depends on time only through the nuclear positions, we also write $\mathbf{S}(t)\equiv\mathbf{S}(\boldsymbol{R}(t))$ below. The elements $\braket{\phi_\mu | \nabla_A \phi_\nu}$ have a form similar to that of the nonadiabatic coupling vectors between many-electron states, but here describe purely geometric basis transport. In particular, they account for the change in the orbital expansion as the basis functions move with the nuclei.

Equation~\ref{eq:EOM_wf} cannot be integrated directly, for two reasons. First, the Hamiltonian $\mathbf{H}[\rho(t)]$ is nonlinear through its dependence on the instantaneous density and hence on the coefficients themselves. Second, the matrices $\mathbf{D}$ and $\mathbf{S}$ carry the geometric dependence of a basis that is itself in motion, which makes stable, unitary propagation delicate.\cite{artacho2017quantum}

We therefore apply a second-order symmetric Trotter (Strang) splitting of the short-time propagator %
\begin{equation}\label{eq:split_evolution}
     \hat{U}(\Delta t) \simeq\hat{U}_{\mathrm{orb}}\!\left(\frac{\Delta t}{2}\right) \hat{U}_{\mathrm{coeff}}(\Delta t)\, \hat{U}_{\mathrm{orb}}\!\left(\frac{\Delta t}{2}\right)
\end{equation}
Here $\hat{U}_{\mathrm{orb}}$ transports the orbitals along the trajectory from one geometry to another time $\Delta t/2$ later, %
whereas $\hat{U}_{\mathrm{coeff}}$ evolves the TDKS coefficients forward by time $\Delta t$ at the fixed midpoint geometry.
Splitting the propagator this way replaces the coupled evolution by two sub-problems that can each be integrated accurately. %

The first operation, $\hat{U}_{\mathrm{orb}}$, corresponds to the $\mathbf{D}$ term of Eq.~\ref{eq:EOM_wf}. Rather than evaluating the derivative-coupling matrix explicitly,\cite{kolesov2016real} we use the numerically more stable L\"owdin-orthonormalized transport scheme,\cite{tomfohr2001time,artacho2017quantum} which maps coefficients expressed in the basis at geometry $\boldsymbol{R}$ onto the basis at $\boldsymbol{R}'$,
\begin{equation}\label{eq:transport}
     \tilde{\mathbf{C}} = \mathbf{S}(\boldsymbol{R}')^{-\frac{1}{2}} \mathbf{S}(\boldsymbol{R})^{\frac{1}{2}} \mathbf{C}
\end{equation}
where $\mathbf{C}$ is a matrix of AO coefficients $c_{\mu n}$ of the TDKS orbitals and $\tilde{\mathbf{C}}$ contains the corresponding coefficients in the basis at $\boldsymbol{R}'$. The transport is applied twice per step: first, from $\boldsymbol{R}(t)$ to ${\boldsymbol{R}(t+\Delta t/2)}$, which turns $\mathbf{C}(t)$ into $\tilde{\mathbf{C}}(t)$; then, after the coefficients are updated from $\tilde{\mathbf{C}}(t)$ to $\tilde{\mathbf{C}}(t+\Delta t)=\hat{U}_\mathrm{coeff}\tilde{\mathbf{C}}(t)$, the basis is transported from $\boldsymbol{R}(t+\Delta t/2)$ to $\boldsymbol{R}(t+\Delta t)$, which turns $\tilde{\mathbf{C}}(t+\Delta t)$ into $\mathbf{C}(t+\Delta t)$.
Because the transformation is a product of matrix square roots of overlap matrices, it preserves orthonormality exactly: $\tilde{\mathbf{C}}^\dagger\mathbf{S}(\boldsymbol{R}')\tilde{\mathbf{C}}={\mathbf{C}}^\dagger\mathbf{S}(\boldsymbol{R}){\mathbf{C}}=\mathbf{I}$ (using the fact that $\mathbf{S}$ is a Hermitian matrix). It is nevertheless an approximation to the exact action of the $\mathbf{D}$ term: it neglects the part of the basis motion that is not reflected in the change of the overlap matrix, for example a rigid translation of the molecule, which leaves $\mathbf{S}$ unchanged.\cite{artacho2017quantum} A projection using the mixed overlap $\langle\phi_\mu(\boldsymbol{R}')|\phi_\nu(\boldsymbol{R})\rangle$ would retain this part but does not in general preserve orthonormality in a finite basis.

The second operation, $\hat{U}_{\mathrm{coeff}}$, evolves the electronic coefficients and corresponds to the $\mathbf{H}$ term of Eq.~\ref{eq:EOM_wf}. In the transported basis, the coefficients then obey
\begin{equation}
    \mathrm{i} \tilde{\mathbf{S}} \dot{\tilde{\mathbf{C}}} = \alpha \mathbf{H}[\rho] \tilde{\mathbf{C}}
\end{equation}
with $\tilde{\mathbf{S}} \equiv \mathbf{S}(\boldsymbol{R}(t+\Delta t/2))$ and with $\mathbf{H}$ constructed at the same geometry. We integrate this equation with the implicit Crank--Nicolson scheme,\cite{crank1947practical,castro2004propagators} chosen because it is unitary and time reversible,
\begin{equation} \label{eq:CN_scheme}
  \left[ \tilde{\mathbf{S}} + \mathrm{i}\alpha \frac{\Delta t}{2} \bar{\mathbf{H}} \right] \tilde{\mathbf{C}}(t+\Delta t) =
  \left[ \tilde{\mathbf{S}} - \mathrm{i}\alpha \frac{\Delta t}{2} \bar{\mathbf{H}} \right] \tilde{\mathbf{C}}(t)
\end{equation}
where the time-averaged Hamiltonian is
\begin{equation}
    \bar{\mathbf{H}} = \frac{1}{2} \left( \mathbf{H}[\rho(t)] + \mathbf{H}[\rho(t+\Delta t)] \right)
\end{equation}
Because $\mathbf{H}[\rho(t+\Delta t)]$ depends on a final density that is not known beforehand, Eq.~\ref{eq:CN_scheme} is nonlinear and must be solved self-consistently. We initialize the time-averaged Hamiltonian from its value at the start of the step, solve Eq.~\ref{eq:CN_scheme}, rebuild $\bar{\mathbf{H}}$ from the resulting density, and repeat.\cite{zhu2018self} Without acceleration, this fixed-point iteration converges slowly in the stiff, large-$\alpha$ regime, so we accelerate it with Anderson mixing\cite{anderson1965iterative} over the history of Hamiltonian matrices, which constructs each new input Hamiltonian as the linear combination of previous input and output Hamiltonians that minimizes the norm of their residual, and stop when the largest change in the molecular-orbital coefficients between successive iterations falls below the convergence threshold. Because consecutive geometries along the trajectory are close, the initial guess is good and convergence is typically reached within a few iterations.

The nuclear force for each state $N$ is the analytic gradient of its potential energy $V_N$, evaluated with the instantaneous propagated density, as in Ehrenfest dynamics. Writing $\mathbf{P}=\sum_n f_n \mathbf{c}_n\mathbf{c}_n^{\dagger}$ for the one-particle density matrix in the atomic-orbital basis, built from the columns $\mathbf{c}_n$ of $\mathbf{C}$ with the occupations of Sec.~\ref{sec:target_states}, the force on nucleus $A$ for this state is
\begin{equation}\label{eq:ehrenfest_force}
    \boldsymbol{F}_A = -\left.\frac{\partial V_N}{\partial \boldsymbol{R}_A}\right|_{\mathbf{C}}
    + \operatorname{Tr}\!\left[\frac{\partial \mathbf{S}}{\partial \boldsymbol{R}_A}
    \,\frac{1}{2}\left( \mathbf{S}^{-1}\mathbf{H}\mathbf{P} + \mathbf{P}\mathbf{H}\mathbf{S}^{-1} \right)\right]
\end{equation}
The first term is the explicit derivative of $V_N$ at fixed AO coefficients. It contains the Hellmann--Feynman force,\cite{feynman1939forces} including the nuclear repulsion, together with derivatives of the AO integrals that arise because the basis functions travel with the nuclei. The second term comes from the derivative of the overlap matrix; together with these integral derivatives it constitutes the Pulay force.\cite{pulay1969ab} In Born--Oppenheimer dynamics this term involves the energy-weighted density matrix \cite{MarxHutterBook} $\mathbf{W}=\sum_n f_n \varepsilon_n \mathbf{c}_n\mathbf{c}_n^{\dagger}$,  whose very definition presupposes that the orbitals are eigenstates of the Kohn--Sham Hamiltonian with eigenvalues $\varepsilon_n$. Propagated TDKS orbitals are not in general eigenstates of $\hat{H}_{\mathrm{KS}}$, so no such eigenvalues exist, and $\tfrac12(\mathbf{S}^{-1}\mathbf{H}\mathbf{P}+\mathbf{P}\mathbf{H}\mathbf{S}^{-1})$ takes the place of $\mathbf{W}$\cite{li2005ab, ojanpera2012nonadiabatic,xu2024lagrangian}; the two coincide whenever the orbitals do satisfy the stationary Kohn--Sham equations, so Eq.~\ref{eq:ehrenfest_force} reduces to the familiar Born--Oppenheimer gradient in that limit. Following common practice,\cite{li2005ab,ojanpera2012nonadiabatic} we omit the further terms involving the velocities of the basis functions, which contribute to momentum conservation but not to the energy gradient.

Algorithm~\ref{alg:propagation} summarizes the propagation scheme. It applies equally to the ground- and the excited-state TDKS orbitals: each is initialized from a converged static SCF solution at the starting geometry and then propagated independently along the nuclear trajectory $\boldsymbol{R}(t)$. Provided $\alpha$ is large enough for the propagated states to follow their adiabatic states (Sec.~\ref{sec:following_theory}), each propagated state reproduces the energy and force of the corresponding adiabatic potential-energy surface on the fly, with neither repeated diagonalization nor explicit optimization of the energy functional at every step. This already allows Born--Oppenheimer dynamics on either the ground or the excited state. To treat nonadiabatic processes we still need a surface-hopping scheme that determines which state drives the nuclei, together with the couplings between the two states on which that scheme relies. This is the subject of the next section.

\begin{pseudocode}{Propagation of a single TDKS state over one step, from $\boldsymbol{R}(t)$ to $\boldsymbol{R}(t+\Delta t)$.}
\label{alg:propagation}
\ttfamily\frenchspacing\raggedright\noindent
\textbf{input:} coefficients $\mathbf{C}\equiv\mathbf{C}(t)$ at $\boldsymbol{R}(t)$, scaling $\alpha$, step $\Delta t$; \ $\tilde{\mathbf{S}}\equiv\mathbf{S}(t+\Delta t/2)$ \\[5pt]
transport to the midpoint basis: $\tilde{\mathbf{C}} \leftarrow \tilde{\mathbf{S}}^{-1/2}\,\mathbf{S}(t)^{1/2}\,\mathbf{C}$ \\[3pt]
initialize $\mathbf{H}' \leftarrow \mathbf{H}[\rho(\tilde{\mathbf{C}})]$ and $\bar{\mathbf{H}} \leftarrow \mathbf{H}'$ at $\boldsymbol{R}(t+\Delta t/2)$ \\[3pt]
\textbf{repeat} \\
\hspace*{2em}solve $(\tilde{\mathbf{S}}+\mathrm{i}\alpha\Delta t\,\bar{\mathbf{H}}/2)\,\tilde{\mathbf{C}}' = (\tilde{\mathbf{S}}-\mathrm{i}\alpha\Delta t\,\bar{\mathbf{H}}/2)\,\tilde{\mathbf{C}}$ for $\tilde{\mathbf{C}}'$ \\
\hspace*{2em}update $\mathbf{H}'$ by Anderson mixing with $\mathbf{H}[\rho(\tilde{\mathbf{C}}')]$; \ $\bar{\mathbf{H}} \leftarrow (\mathbf{H}[\rho(\tilde{\mathbf{C}})]+\mathbf{H}')/2$ \\
\textbf{until} the change in $\tilde{\mathbf{C}}'$ is below the threshold \\[3pt]
transport to the end-point basis: $\mathbf{C}' \leftarrow \mathbf{S}(t+\Delta t)^{-1/2}\,\tilde{\mathbf{S}}^{1/2}\,\tilde{\mathbf{C}}'$ \\[5pt]
\textbf{return} $\mathbf{C}'\equiv\mathbf{C}(t+\Delta t)$, with energy and forces at $\boldsymbol{R}(t+\Delta t)$ evaluated from $\rho(\mathbf{C}')$
\end{pseudocode}

\section{Surface Hopping in a Complex Adiabatic Basis}\label{sec:sh}

To carry out nonadiabatic molecular dynamics, we combine the on-the-fly real-time propagation scheme developed in Sec.~\ref{sec:rttddft} with the mapping approach to surface hopping (MASH),\cite{MASH,MASHreview} which provides equations of motion for the time-dependent coefficients of the two adiabatic electronic states and a rule for when the nuclei switch surface. The standard MASH formulation shares much of its algorithmic structure with FSSH but is deterministic and time reversible.\cite{MASHEOM} It is free of the internal-consistency problems of FSSH,\cite{Granucci2007FSSH,Subotnik2016review} correctly obeys detailed balance,\cite{thermalization} and can be used to calculate reaction rates,\cite{MASHrates} sample rare events,\cite{ghamari2026nonadiabatic,Dellago2026MASHTPS} capture electronic coherences,\cite{Mannouch2024coherence,MASHcoh} describe open quantum systems,\cite{asnaashari2026open} and perform on-the-fly simulations of photochemistry.\cite{mannouch2024quantum,cyclobutanone,Hutton2024cyclobutatone}

Although MASH was originally formulated for two states, various multistate generalizations have been proposed.\cite{unSMASH,Runeson2023MASH,Granuuci_multiMASH} Using MASH in the present framework requires two extensions. First, previous surface-hopping implementations based on static electronic-structure calculations assume real-valued wavefunctions,\cite{doltsinis2002nonadiabatic,tapavicza2007trajectory,pradhan2018non,malis2020trajectory,MASHEOM,cyclobutanone,mannouch2024quantum} whereas real-time propagation necessarily yields complex-valued states carrying time-dependent phases; we therefore recast the MASH equations of motion in a complex adiabatic basis. Second, the ground and excited states are obtained from different SCF solutions, and so do not share a common KS Hamiltonian and are not mutually orthogonal; we therefore evaluate the time-derivative couplings and nonadiabatic coupling vectors numerically between the two independently propagated adiabatic states.

\subsection{MASH Formulation for Complex Adiabatic States}\label{sec:mash_complex}

First, we assume that the propagated adiabatic states have been orthogonalized using the procedure that we will describe in Sec.~\ref{sec:dto}. The time-dependent electronic wavefunction can then be written as $|\psi (t)\rangle=c_0(t)|\Phi_0(\boldsymbol{R}(t))\rangle+c_1 (t)|\Phi_1 (\boldsymbol{R}(t))\rangle$, so that $c_0$ and $c_1$ describe the amplitude on each state. In MASH, this two-state amplitude is mapped onto a classical spin vector $\vec{S}=\{S_x,S_y,S_z\}$ on the unit Bloch sphere, whose Cartesian components are the expectation values of the Pauli operators, $S_\mu=\langle\hat{\sigma}_\mu\rangle$.\cite{MASH} In the two-state adiabatic basis $\{|\Phi_0\rangle, |\Phi_1\rangle\}$ these operators read
\begin{subequations}
\begin{align}
    \hat{\sigma}_x &= |\Phi_0\rangle\langle \Phi_1| + |\Phi_1 \rangle\langle \Phi_0| \\
    \hat{\sigma}_y &= i(|\Phi_0\rangle\langle \Phi_1| - |\Phi_1 \rangle\langle \Phi_0|) \\
    \hat{\sigma}_z &= |\Phi_1 \rangle\langle \Phi_1| - |\Phi_0 \rangle\langle \Phi_0|
\end{align}
\end{subequations}
Hence, the mapping variables are $S_x=2\operatorname{Re}(c_0^{*}c_1)$, $S_y=-2\operatorname{Im}(c_0^{*}c_1)$ and $S_z=|c_1|^2-|c_0|^2$.

To obtain the spin dynamics, we first write the equations of motion for the coefficients in the complex adiabatic basis. KS-DFT does not provide a universal many-electron Hamiltonian common to these state-specific solutions, so we instead define an effective diagonal Hamiltonian whose eigenvalues are the adiabatic potential-energy surfaces $\{V_N(\boldsymbol{R})\}$, where $V_N(\boldsymbol{R})$ denotes the unscaled potential energy $V(\boldsymbol{R})$ of Sec.~\ref{sec:variational}, evaluated for state $N$ with the occupations specified in Sec.~\ref{sec:target_states}.
The states are coupled by the time-derivative coupling matrix $T_{MN} = \braket{\Phi_M | \partial_t \Phi_N}$, which is anti-Hermitian, with purely imaginary diagonal elements that need not vanish.
In this basis, the coefficients $\{c_0, c_1\}$ obey the time-dependent Schr\"odinger equation,
\begin{equation}\label{eq:coeff_eom}
    \dot{c}_M = -\mathrm{i}V_M c_M - \sum_N T_{MN} c_N
\end{equation}
which, together with $T_{NN}^{*}=-T_{NN}$ and $T_{10}=-T_{01}^{*}$, gives the equation of motion of the spin vector on the Bloch sphere,
\begin{equation} \label{eq:spin_dynamics_final}
    \begin{pmatrix} \dot{S}_x \\ \dot{S}_y \\ \dot{S}_z \end{pmatrix} =
    \begin{pmatrix}
    0 & -\Omega & -2\operatorname{Re}(T_{01}) \\
    \Omega & 0 & -2\operatorname{Im}(T_{01}) \\
    2\operatorname{Re}(T_{01}) & 2\operatorname{Im}(T_{01}) & 0
    \end{pmatrix}
    \begin{pmatrix} S_x \\ S_y \\ S_z \end{pmatrix}
\end{equation}
where $\Omega=\Delta V-\operatorname{Im}(T_{00}-T_{11})$ and $\Delta V=V_1-V_0$.

The diagonal couplings $T_{NN}$ vanish for real-valued adiabatic states, whereas in a complex basis they account for the phase evolution of the basis states. In Eq.~\ref{eq:spin_dynamics_final}, they enter through $\Omega$, which sets the precession rate about the $z$ axis in the chosen phase convention. For a model Hamiltonian with eigenenergies $E_N=V_N$, the natural phase evolution of $|\Phi_N(t)\rangle=e^{-\mathrm{i}E_Nt}|\Phi_N\rangle$ gives $T_{NN}=-\mathrm{i}E_N$ and $\Omega=0$. The basis states then carry the dynamical phases, and the coefficients are expressed in a co-rotating frame. %

However, under real-time KS propagation, the phases carried by the basis states need not be generated by the total energies $V_N$. For example, at fixed nuclear geometry and before any additional gauge fixing, a determinant built from stationary orbitals acquires a phase governed by the sum of the occupied orbital energies multiplied by $\alpha$ from Eq.~\ref{eq:tdks_eom}.  In KS-DFT, the sum of occupied orbital energies generally differs from the total energy $V_N$. In this case, the diagonal terms $T_{NN}$ account for the phase evolution of the basis in Eq.~\ref{eq:coeff_eom} and cannot be set to zero.

The dynamics is formally independent of the phase convention due to gauge invariance.
To make this explicit, consider a change of the basis functions by an arbitrary real phase $\theta_N(t)$, accompanied by a corresponding change of the expansion coefficients,
\[
    |\Phi_N'\rangle=e^{\mathrm{i}\theta_N(t)}|\Phi_N\rangle,
    \qquad
    c_N'=e^{-\mathrm{i}\theta_N(t)}c_N,
\]
such that the total electronic wavefunction is unchanged. Under this transformation, $T_{01}$ and $S_x-\mathrm{i}S_y$ acquire opposite phase factors, leaving the population equation
\begin{equation}\label{eq:secular}
    \dot{S}_z = 2\operatorname{Re}(T_{01})S_x + 2\operatorname{Im}(T_{01})S_y = 2\operatorname{Re}\!\left[T_{01}\left(S_x-\mathrm{i}S_y\right)\right]
\end{equation}
unchanged. The momentum-rescaling combination $\operatorname{Re}[\boldsymbol{d}_{01}(S_x-\mathrm{i}S_y)]$ in Eq.~\ref{eq:mash_force_final} is invariant for the same reason. Thus, populations, hopping events and rescaling directions are independent of the phase convention used for the basis states.

The population equation also influences the nuclear dynamics, since crossing the Bloch-sphere equator at $S_z=0$ triggers an attempt to switch the active surface. The nuclear equations of motion follow from requiring conservation of the total MASH energy,\cite{MASH} $\mathcal{E} = \sum_{A}\boldsymbol{P}_A^2/2M_A + h(-S_z)V_0 + h(S_z)V_1$, where $h$ is the Heaviside step function and $M_A$ is the mass of atom $A$. Setting $\mathrm{d}\mathcal{E}/\mathrm{d}t=0$, substituting $\dot{S}_z$ from Eq.~\ref{eq:secular} and using $\dot{\boldsymbol{R}}_A=\boldsymbol{P}_A/M_A$ and $T_{01}=\sum_A\dot{\boldsymbol{R}}_A\cdot\boldsymbol{d}^A_{01}$, where $\boldsymbol{d}^A_{01}=\langle\Phi_0|\nabla_A\Phi_1\rangle$ is the nonadiabatic coupling vector (NACV) between the two propagated states, gives the generalized nuclear force
\begin{equation}\label{eq:mash_force_final}
    \dot{\boldsymbol{P}}_A = -\nabla_A V_0 h(-S_z) - \nabla_A V_1 h(S_z) - 2\Delta V \delta(S_z) \left[ \operatorname{Re}(\boldsymbol{d}^{A}_{01}) S_x + \operatorname{Im}(\boldsymbol{d}^{A}_{01}) S_y \right]
\end{equation}

The impulsive term at $S_z=0$ involves the NACV $\boldsymbol{d}_{01}$. For complex-valued states, only the \emph{direction} of the combination $\operatorname{Re}(\boldsymbol{d}^{A}_{01}) S_x + \operatorname{Im}(\boldsymbol{d}^{A}_{01}) S_y$ enters the momentum rescaling. Each time a hop is attempted, the kinetic energy available along that direction is compared with the energy gap to decide whether the hop is energetically allowed; if it is, the momentum is rescaled along that direction, and if not the hop is frustrated. Standard MASH handles a frustrated hop by reversing the momentum along the NACV.\cite{MASH,MASHreview} We instead use the recently proposed $\pi$-MASH prescription to flip the spin vector, $\vec{S}\leftarrow-\vec{S}$, and leave the momentum unchanged.\cite{furlanetto2026frustrated}

\subsection{DTO Representation of the Adiabatic States}\label{sec:dto}

Evaluating the time-derivative coupling matrix $\mathbf{T}$ and the NACV $\boldsymbol{d}_{01}$ requires a well-defined representation of the adiabatic states. Within the KS-DFT framework, constructing many-electron wavefunctions from KS Slater determinants is a practical approximation routinely used in static $\Delta$SCF.\cite{doltsinis2002nonadiabatic,pradhan2018non} Even in static $\Delta$SCF, the ground and excited states are optimized independently, so their mutual orthogonality is not enforced. Real-time propagation introduces an additional ambiguity: the two equally occupied active spatial orbitals of the excited state may mix unitarily without changing the KS density, while changing the spin-adapted excited-state wavefunction constructed from them.
In the strong-coupling region, this mixing can give both active orbitals substantial components in the occupied ground-state space. The singlet constructed from them can then develop a large overlap with the ground-state determinant (Sec.~\ref{sec:dto_results}, Fig.~\ref{fig:DTO_overlap}), compromising the overlap-based evaluation of the couplings.
This behavior resembles the orbital-rotation instability found in ROKS simulations.\cite{kowalczyk2013excitation,buchel2024simulation}

We fix the particle--hole representation by transforming the active TDKS orbitals into dynamical transition orbitals (DTOs).\cite{zhou2021dynamical} The construction uses spin-restricted orbitals with equal occupations in the active space. We first build the projector onto the occupied ground-state ($N=0$) orbitals,
\begin{equation}\label{eq:dto_projector}
    \hat{P}_0(t) = \sum_{n \in \text{occ}} \ket{\psi_{n}^{(0)}(t)} \bra{\psi_{n}^{(0)}(t)}
\end{equation}
and then the projection matrix $\mathbf{M}(t)$ within the subspace spanned by the two time-dependent singly occupied orbitals of the excited state ($N=1$), $\{\psi_{a}^{(1)},\psi_{b}^{(1)}\}$,
\begin{equation}\label{eq:dto_M}
    M_{ij}(t) = \braket{\psi_{i}^{(1)}(t) | \hat{P}_0(t) | \psi_{j}^{(1)}(t)}, \quad i,j \in \{a, b\}
\end{equation}
The eigenvalues of $\mathbf{M}(t)$ are the weights with which the corresponding orbital
combinations project onto the occupied ground-state space, analogous to
natural transition orbitals.\cite{li2011time,martin2003natural} The eigenvectors
define linear combinations of the two active orbitals: one corresponding to an eigenvalue
close to unity identifies a ``hole'' orbital lying almost entirely in the occupied
ground-state space, and the other with an eigenvalue close to zero identifies a
``particle'' orbital nearly orthogonal to that space. We use the hole and particle
DTOs to construct the excited-state singlet $\Phi_1$ according to Eq.~\ref{eq:SD_es}. When the particle DTO has little weight in the occupied ground-state space, $\Phi_1$ has a small overlap with the ground state $\Phi_0$, although exact orthogonality is not imposed. The overlap matrix between these states at
successive time steps (Eq.~\ref{eq:U_def}) is then used to evaluate the time-derivative
coupling matrix $\mathbf{T}$, as described in Sec.~\ref{sec:nac}.

The projector $\hat{P}_0$ depends only on the occupied \emph{subspace} of $\Phi_0$, so the DTO basis is insensitive to unitary rotations among the occupied ground-state orbitals. Within the excited-state active space, the equal occupations $f_a=f_b=1$ ensure that the density $\rho=\sum_n f_n|\psi_n|^2$ is invariant under unitary mixing of $\{a,b\}$. The DTO transformation therefore leaves the density, energy and nuclear forces unchanged, while fixing the constructed many-electron wavefunctions used for overlaps. Explicit rotation is needed only in the active space, since a unitary rotation among doubly occupied orbitals changes a Slater determinant only by an overall phase.

After the particle--hole directions have been selected, each DTO still has an arbitrary phase. Independent diagonalization of Eq.~\ref{eq:dto_M} at each step can thus produce discontinuous phase changes and artificial spikes in the overlap-based couplings. We therefore align each DTO with the corresponding orbital at the previous step by choosing its phase so that their overlap is real and positive, provided the overlap is nonzero. This maintains a continuous gauge along the trajectory.

\subsection{Numerical Evaluation of Nonadiabatic Couplings}\label{sec:nac}
The time-derivative coupling matrix $\mathbf{T}$ is needed at every step to propagate the spin-mapping variables according to Eq.~\ref{eq:spin_dynamics_final}. Since the many-electron wavefunctions constructed in the DTO gauge are expressed in terms of Slater determinants, their overlaps between successive time steps are directly available, as in overlap-based estimates of the time-derivative coupling.\cite{hammes1994proton} We obtain $\mathbf{T}$ from these overlaps using local diabatization.\cite{granucci2001direct,plasser2012surface,meek2014evaluation,MASHEOM}

In local diabatization, the adiabatic basis at time $t$ is held fixed and used as a diabatic basis over a short interval, with the dynamics approximated within this two-state space. The change between the basis functions at $t$ and $t+\Delta t$, including the phases carried by the propagated states, is measured by the overlap matrix
\begin{equation}\label{eq:U_def}
    U_{MN} = \braket{\Phi_M(\boldsymbol{R}(t)) | \Phi_N(\boldsymbol{R}(t+\Delta t))}
\end{equation}
The overlap of two Slater determinants is the determinant of their occupied
spin-orbital overlap matrix, which factorizes into one spatial-orbital determinant
per spin channel. Applying this rule to Eqs.~\ref{eq:SD_gs} and \ref{eq:SD_es}
(see Fig.~\ref{fig:mo_diagrams}) gives
\cite{doltsinis2002nonadiabatic,hait2016prediction}
\begin{equation}\label{eq:U_det}
    \mathbf{U} =
    \begin{pmatrix}
        \mathcal{D}_{gg}^{2} & \sqrt{2}\,\mathcal{D}_{gh}\mathcal{D}_{gl} \\[3pt]
        \sqrt{2}\,\mathcal{D}_{hg}\mathcal{D}_{lg} & \mathcal{D}_{hh}\mathcal{D}_{ll} + \mathcal{D}_{hl}\mathcal{D}_{lh}
    \end{pmatrix}
\end{equation}
Here $g$ denotes the occupied ground-state spatial orbitals; $h$ and $l$ contain the doubly occupied excited-state orbitals plus the hole and particle DTO, respectively. Each $\mathcal{D}$ is the determinant of a spatial-orbital overlap matrix, with the first subscript referring to $t$ and the second to $t+\Delta t$. For example, $\mathcal{D}_{gh}=\det\mathbf{O}_{gh}$ with
\begin{equation}\label{eq:O_ao}
    \mathbf{O}_{gh}=\mathbf{C}_g^{\dagger}(t)\,\mathbf{S}\big(\boldsymbol{R}(t),\boldsymbol{R}(t+\Delta t)\big)\,\mathbf{C}_h(t+\Delta t)
\end{equation}
where $\mathbf{C}_g$ and $\mathbf{C}_h$ contain the corresponding AO coefficients, and $S_{\mu\nu}(\boldsymbol{R},\boldsymbol{R}')=\braket{\phi_\mu(\boldsymbol{R})|\phi_\nu(\boldsymbol{R}')}$ is the AO overlap matrix between the two geometries. The remaining orbital overlap matrices are constructed analogously.

As the nuclear geometry changes, the states at $t+\Delta t$ may acquire components outside the space spanned by the two states at $t$, so their $2\times2$ overlap matrix need not be unitary. In the trajectories reported below, this deviation from unitarity appears as singular values $10^{-3}$--$10^{-4}$ below unity per step. Although small at each step, these deviations can accumulate over a trajectory. Using a non-unitary $\mathbf{U}$ in the matrix logarithm introduced below would give $\bar{\mathbf{T}}$ a Hermitian component, inconsistent with the anti-Hermitian couplings required for the norm-conserving two-state evolution underlying Eq.~\ref{eq:spin_dynamics_final}. We therefore project $\mathbf{U}$ onto the nearest unitary matrix through its singular-value decomposition,\cite{jain2016efficient}
\begin{equation}\label{eq:U_corr}
    \mathbf{U} = \mathbf{W} \boldsymbol{\Sigma} \mathbf{Z}^\dagger ,
    \qquad
    \mathbf{U}_{\text{corr}} = \mathbf{W} \mathbf{Z}^\dagger
\end{equation}
where $\mathbf{W}$ and $\mathbf{Z}$ are unitary and $\boldsymbol{\Sigma}$ holds the singular values.

Since $\mathbf{U}_{\text{corr}}$ is unitary, its generator is the anti-Hermitian coupling matrix $\bar{\mathbf{T}}$, provided the electronic states vary smoothly, such that $\mathbf{U}_{\text{corr}}\approx\exp(\bar{\mathbf{T}}\Delta t)$ over a short step, and we can obtain $\bar{\mathbf{T}}$ from\cite{jain2016efficient}
\begin{equation} \label{eq:matrix_logm_final}
    \bar{\mathbf{T}} \approx \frac{\operatorname{logm}\left(\mathbf{U}_{\text{corr}}\right)}{\Delta t}
\end{equation}
where $\operatorname{logm}$ denotes the principal matrix logarithm. This gives the full coupling matrix, including its diagonal and off-diagonal elements in the complex basis.

Whereas $\bar{\mathbf{T}}$ is required at every step, the NACV $\boldsymbol{d}_{01}$ is evaluated only at an attempted hop, to determine whether the hop is energetically allowed and, if so, to rescale the momentum.
Equation~\ref{eq:matrix_logm_final} uses the overlap logarithm to estimate how the electronic basis changes per unit time. The NACV describes the corresponding change per unit nuclear displacement. This correspondence motivates estimating the NACV from the spatial variation of the overlap logarithm, with the initial states and propagation interval held fixed. We evaluate it within the same local-diabatization framework, in which $\mathbf{U}_{\mathrm{corr}}(\boldsymbol{R},\boldsymbol{R}';\Delta t)$ denotes the unitarized overlap between states at $\boldsymbol{R}$ and those propagated to $\boldsymbol{R}'$ in a time $\Delta t$ by Eq.~\ref{eq:tdks_eom}. The NACV is then estimated as
\begin{equation}
    \boldsymbol{d}_{01}^{A}(\boldsymbol{R}) \approx \left.\nabla'_A \left[\operatorname{logm}\left(\mathbf{U}_{\mathrm{corr}}(\boldsymbol{R},\boldsymbol{R}';\Delta t)\right)\right]_{01}\right|_{\boldsymbol{R}'=\boldsymbol{R}} \label{eq:nacv_fd}
\end{equation}
Here $\nabla'_A$ denotes the gradient with respect to $\boldsymbol{R}'_A$, with the initial states and $\Delta t$ held fixed. Numerically, this gradient is evaluated by central finite differences using geometries displaced by $\pm\Delta R$ along each nuclear Cartesian coordinate, with the overlaps unitarized as in Eq.~\ref{eq:U_corr}. The propagation interval $\Delta t$ used here is examined in Sec.~\ref{sec:nac_results} and need not equal the trajectory time step.

According to the quantum adiabatic principle (Sec.~\ref{sec:following_theory}), the electronic states need sufficient propagation time to follow their adiabatic states across the displacement. %
The extraction therefore requires both a sufficiently small $\Delta R$ for the central difference and a sufficiently long propagation interval, or equivalently a large enough $\alpha$ by the reparametrization of Eq.~\ref{eq:tau_reparam}.

The extracted $\boldsymbol{d}_{01}$ is complex because the constructed many-electron wavefunctions carry real-time phases.  However, we know that in principle there exists a basis in which the instantaneous adiabatic states may be chosen real, with an underlying real coupling $\bar{\boldsymbol{d}}_{01}$. To first order in $\Delta R$ and within the two-state space, propagation across the displacement multiplies all Cartesian components by the same complex factor $z=|z|e^{\mathrm{i}\varphi}$, which depends only on $\alpha\,\Delta E\,\Delta t$. Writing $\boldsymbol{d}_{01}=z\,\bar{\boldsymbol{d}}_{01}$, the combination entering the rescaling force of Eq.~\ref{eq:mash_force_final} is then
\begin{equation}\label{eq:rescale_direction}
    \operatorname{Re}(\boldsymbol{d}_{01})S_x+\operatorname{Im}(\boldsymbol{d}_{01})S_y = |z|\,\bar{\boldsymbol{d}}_{01}\left(S_x\cos\varphi + S_y\sin\varphi\right)
\end{equation}
For a nonzero scalar prefactor, this vector is parallel to $\bar{\boldsymbol{d}}_{01}$, so $z$ changes its magnitude but not its axis within this approximation. No additional phase fixing is required. In the present implementation, we use $\operatorname{Re}(\boldsymbol{d}_{01})$ alone, which is parallel to $\bar{\boldsymbol{d}}_{01}$ provided $\operatorname{Re}z\neq0$.  Note that the magnitude and sign of the NACV are irrelevant for the momentum rescaling and that only the direction is required.

Each such NACV evaluation for a system of $N_{\mathrm{atom}}$ atoms costs $6N_{\mathrm{atom}}$ additional short propagations---the most expensive step of the scheme---but it is only needed at attempted hops, which occur only a few times per trajectory. Nevertheless, analytical NACVs between the propagated states, along the lines available for other mean-field excited-state methods,\cite{billeter2006calculation,ramos2021nonadiabatic} would remove this finite-difference cost and are planned for future work.

\subsection{Integration of the Coupled Dynamics}\label{sec:integration}

Algorithm~\ref{alg:mash} uses the propagated electronic states and the numerically evaluated couplings of Sec.~\ref{sec:nac} to evolve the coupled two-state MASH dynamics. The nuclei are integrated with velocity Verlet, while the two electronic states are propagated independently at the midpoint geometry of each step and transported to its end using Algorithm~\ref{alg:propagation}. Their state-specific energies and time-derivative coupling matrix drive the spin evolution, while $\operatorname{sign}(S_z)$ selects the active surface that supplies the nuclear force. Here $\boldsymbol{F}_A=-\nabla_A V_a$ denotes the smooth force on that surface, whereas the impulsive contribution in Eq.~\ref{eq:mash_force_final} is handled by momentum rescaling at a hop. The implementation uses the nonreversible local-diabatization scheme of Ref.~\citenum{MASHEOM}. A hop attempt is assessed at the end of the step, after both nuclear half kicks; the second half kick uses the force of the state active at the start of the step, and the active state and force passed to the next step are set only after the hop has been accepted or rejected. Updates for steps in which $S_z$ does not change sign are symmetric under time reversal, but this symmetry is not imposed on steps containing a hop attempt. More efficient reversible schemes, which locate the crossing within the step and split the propagation there to apply momentum rescaling before continuing\cite{MASHEOM} are left for future work.

\begin{pseudocode}{Coupled MASH dynamics over one time step $\Delta t$, from $\boldsymbol{R}(t)$ to $\boldsymbol{R}(t+\Delta t)$.}
\label{alg:mash}
\ttfamily\frenchspacing\raggedright\noindent
\textbf{input:} positions $\boldsymbol{R}(t)$, momenta $\boldsymbol{P}(t)$, spin $\vec{S}$, states $\Phi_0,\Phi_1$, force $\boldsymbol{F}(t)$ \\[5pt]
\emph{nuclei (Verlet)} \\
\hspace*{2em}half kick: $\boldsymbol{P}_A \leftarrow \boldsymbol{P}_A + (\Delta t/2)\,\boldsymbol{F}_A$ \\
\hspace*{2em}drift: \ $\boldsymbol{R}_A(t+\Delta t/2) \leftarrow \boldsymbol{R}_A(t) + (\Delta t/2)\,\boldsymbol{P}_A/M_A$ \\
\hspace*{2em}\phantom{drift: \ }$\boldsymbol{R}_A(t+\Delta t) \leftarrow \boldsymbol{R}_A(t) + \Delta t\,\boldsymbol{P}_A/M_A$ \\[3pt]
\emph{electrons} \\
\hspace*{2em}propagate $\Phi_0$, $\Phi_1$ to $\boldsymbol{R}(t+\Delta t)$ by Algorithm~\ref{alg:propagation}; obtain $V_0$, $V_1$ and forces \\[3pt]
\emph{nuclei (Verlet)} \\
\hspace*{2em}half kick: $\boldsymbol{P}_A \leftarrow \boldsymbol{P}_A + (\Delta t/2)\,\boldsymbol{F}_A$, with $\boldsymbol{F}_A = -\nabla_A V_{a}(\boldsymbol{R}(t+\Delta t))$ and $a$ the active state at $t$ \\[3pt]
\emph{spin} \\
\hspace*{2em}$\bar{\mathbf{T}} \leftarrow \operatorname{logm}(\mathbf{U}_{\text{corr}})/\Delta t$ from DTO-gauge overlaps, Eqs.~\ref{eq:U_det}{\normalfont--}\ref{eq:matrix_logm_final} \\
\hspace*{2em}propagate $\vec{S}$ by Eq.~\ref{eq:spin_dynamics_final}; $\operatorname{sign}(S_z)$ selects the candidate state $b$ \\[3pt]
\emph{hop} \\
\hspace*{2em}\textbf{if} $b \neq a$ \textbf{then} evaluate $\boldsymbol{d}_{01}$ by Eq.~\ref{eq:nacv_fd} \\
\hspace*{4em}\textbf{if} energetically allowed \textbf{then} rescale $\boldsymbol{P}$ along $\operatorname{Re}(\boldsymbol{d}_{01})$; $a \leftarrow b$ \\
\hspace*{4em}\textbf{else} $\vec{S} \leftarrow -\vec{S}$ (frustrated hop) \\
\hspace*{2em}$\boldsymbol{F}_A(t+\Delta t) \leftarrow -\nabla_A V_{a}(\boldsymbol{R}(t+\Delta t))$ \\[5pt]
\textbf{return} $\boldsymbol{R}(t+\Delta t)$, $\boldsymbol{P}(t+\Delta t)$, $\vec{S}$, $\Phi_0,\Phi_1$, $\boldsymbol{F}(t+\Delta t)$
\end{pseudocode}

\section{Results and Discussion} \label{sec:results}

\subsection{Computational Details}\label{sec:comp_details}

The method was implemented in an in-house Python package built on PySCF\cite{sun2018pyscf,sun2020recent} (version 2.12.1) and applied to \textit{trans}-diimide (\ce{N2H2}) with the 6-31G* basis set.\cite{ditchfield1971self,hehre1972self,hariharan1973influence} Diimide is the simplest azo compound, and its excited states were characterized early by \textit{ab initio} methods.\cite{vasudevan1975theoretical} It provides a minimal model of the azo unit relevant to azobenzene photoisomerization.\cite{bandara2012photoisomerization,ciminelli2004photoisomerization} Two levels of theory are used throughout: real-time Hartree--Fock (RT-HF) and RT-TDDFT with the exchange-only local-density approximation (LDA)\cite{slater1974quantum}. We chose this functional as a demanding numerical test rather than for its accuracy. It gives small energy gaps in the strong-coupling region, where static SCF becomes unreliable and the dynamical propagation is most useful. RT-HF is the complementary case. Its static solutions converge reliably along most of the path and thus provide a reference against which the propagated energies are checked. In both cases, the electronic trajectories are initialized from converged spin-restricted SCF solutions for the ground state ($N=0$) and the first excited state ($N=1$), as described in Sec.~\ref{sec:target_states}. For comparison, static references are computed on exactly the same geometries as the propagated results, using the CIAH orbital optimizer for the ground state and DIIS with MOM for the excited state, with a convergence threshold of $10^{-12}$~Ha and at most 500 iterations.

The dynamical propagation of the electronic degrees of freedom uses the Crank--Nicolson scheme of Eq.~\ref{eq:CN_scheme}. Its predictor--corrector iteration is taken as converged when the largest change in the molecular-orbital coefficients between successive iterations falls below $10^{-5}$, and it is accelerated by Anderson mixing.  A larger $\alpha$ does in principle demand a smaller time step for a given accuracy, since the phase accumulated per step grows in proportion to $\alpha$. In practice, the Anderson-accelerated iteration converges reliably up to the largest $\alpha$ used here, and the convergence of the ensemble results with respect to the time step is shown in the Supporting Information. For trajectory propagation, we use equal electronic and nuclear time steps of $0.4$~a.u.\ ($0.0097$~fs) unless otherwise stated for specific validation tests. This step is small enough for the largest $\alpha$ used, and keeping it fixed makes trajectories computed with different $\alpha$ directly comparable. Such a small step is typical of conventional RT-TDDFT propagation\cite{li2020real,xu2024real} and was chosen conservatively for the present proof-of-principle study. More efficient RT-TDDFT schemes with larger propagation intervals are available\cite{jia2018fast,wang2015efficient} and could be adopted in future work.  %

Initial nuclear coordinates and velocities were generated by Wigner sampling\cite{wigner1932quantum,hillery1984distribution} at 300~K, using the harmonic approximation to the ground-state LDA potential-energy surface (details in Sec.~\ref{sec:ensemble}). To examine the electronic propagation and the evaluation of the couplings in isolation, we first use prescribed nuclear paths. Section~\ref{sec:following} defines two paths of $800$~a.u.\ ($19.4$~fs) through the strong-coupling region, one for RT-HF and one for RT-TDDFT; the RT-TDDFT path is
also used in Secs.~\ref{sec:dto_results} and \ref{sec:spin}.
Both paths start from the same snapshot, taken partway along a longer trajectory started from a Wigner-sampled initial condition on the ground state. Its geometry and nuclear velocities are provided in the Supporting Information to ensure the reproducibility of our results. Starting from the snapshot, each path is generated once by $\alpha=1$ Ehrenfest dynamics, in which the nuclei feel the mean-field force of the instantaneous excited-state density,\cite{xu2024lagrangian,zobavc2025ehrenfest} and the midpoint and end-point geometries of every step are stored. Then, for a range of values of $\alpha$, both electronic states are propagated along the stored trajectory without letting the nuclei respond. Any difference between two values of $\alpha$ thus comes from the electronic propagation alone. No hops are attempted, no momenta are rescaled, and the spin-mapping variables of Sec.~\ref{sec:spin} evolve without acting back on the nuclei. The spin dynamics shown there is thus a diagnostic of the electronic evolution rather than a surface-hopping simulation.

For the full coupled dynamics in Sec.~\ref{sec:ensemble}, we use an ensemble of 100 trajectories drawn from the same Wigner distribution, each propagated from $t=0$ for $4000$~a.u.\ ($96.8$~fs), which is long enough to observe the relaxation from the excited state to the ground state. Here the state-specific forces switch between the two surfaces according to the sign of $S_z$, as prescribed by Eq.~\ref{eq:mash_force_final}, and hops and momentum rescaling are active.

\subsection{Numerical Tests of Adiabatic-State Following}\label{sec:following}

\begin{figure}[htbp]
    \centering
    \begin{subfigure}[t]{0.45\textwidth}
        \centering
        \includegraphics[width=\textwidth]{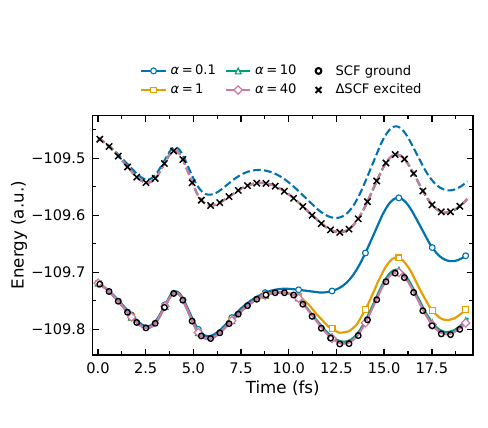}
        \caption{}
    \end{subfigure}
    \begin{subfigure}[t]{0.45\textwidth}
        \centering
        \includegraphics[width=\textwidth]{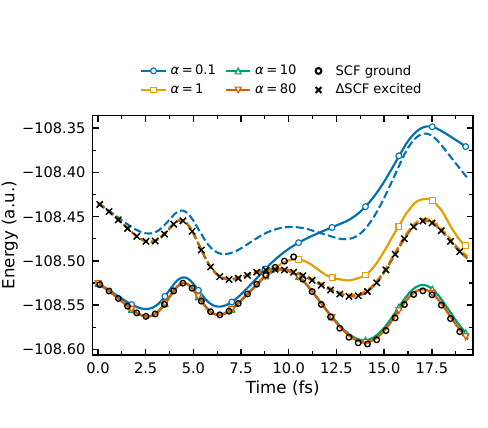}
        \caption{}
    \end{subfigure}
    \caption{Potential-energy profiles along \textit{trans}-diimide nuclear paths obtained with (a) RT-HF and (b) RT-TDDFT. Dynamically propagated results at several values of $\alpha$ are compared with static SCF references. Solid and dashed curves denote the propagated ground- and excited-state energies, respectively; open circles and crosses denote the corresponding static references. Within each panel, all propagated curves and static references are computed on the same nuclear path, generated once by $\alpha=1$ Ehrenfest dynamics.}
    \label{fig:scanning_LDA}
\end{figure}

Figure~\ref{fig:scanning_LDA} compares the dynamically propagated ground- and excited-state energies at several values of $\alpha$ with static SCF references computed along the same nuclear path. This comparison tests adiabatic-state following (Sec.~\ref{sec:following_theory}). We first examine the RT-HF results in Fig.~\ref{fig:scanning_LDA}(a). At $\alpha=0.1$, the electronic response is slow relative to the nuclear motion, thereby narrowing instead of widening the separation of electronic and nuclear frequencies; the propagated wavefunction thus mixes with higher-lying states and departs substantially from the static SCF reference. At $\alpha=1$, the propagated energies initially follow the adiabatic surfaces but depart from them on entering the strong-coupling region near $t=10.4$~fs. As $\alpha$ is increased further, the propagated energies converge systematically to the static solution. In particular, at $\alpha=40$ the timescales are well separated, and both propagated energies agree with the static references along the whole path to within 1\% of the energy gap between the states.

We next repeat the test with RT-TDDFT, a more demanding case because the relevant gaps are considerably smaller than at the RT-HF level. As Fig.~\ref{fig:scanning_LDA}(b) shows, the ground- and excited-state surfaces come close to each other near $t=9.9$~fs. As in the RT-HF case, the propagated energies at $\alpha=0.1$ and 1 depart from the static references. At $\alpha\geq10$ the propagated energies converge towards the static references as $\alpha$ is increased, except in a region where the static ground-state solution leaves the lower branch. In this strong-coupling region standard DIIS failed to converge for every level shift\cite{saunders1973level} we tested; static solutions could be obtained only with the CIAH orbital optimizer, which avoids direct diagonalization of the Fock matrix.

Even where it converges, the static CIAH ground-state solution leaves the lower branch near $t=8.9$~fs and lies up to 0.65~eV above the propagated ground state until about $t=10.6$~fs, where it drops back abruptly, leaving an unphysical kink in the profile. The dynamical result at $\alpha=80$ instead follows a lower-energy, continuous branch through the same region. This suggests that the static solver converges to a higher-lying stationary solution, whereas the propagated wavefunction keeps continuity with the true ground state selected earlier in the trajectory. Since our aim in this work is the dynamical propagation, we have not attempted to fix the static solver to recover the lower branch, although more advanced static optimization schemes\cite{schmerwitz2023calculations} could presumably locate it. The dynamic propagation, by contrast, follows it correctly without any adjustment of constraints or initial guesses. For the excited state, the energy deviation from the static reference is typically below 2~meV over the whole path for $\alpha\geq10$. For the ground state, the energy deviation after the strong-coupling region decreases from about 2.5~eV at $\alpha=1$ to at most about 50~meV at $\alpha \geq 40$.

In this proof-of-principle demonstration of state following, we have not optimized the integration scheme for efficiency or assessed its performance systematically. Static Born--Oppenheimer dynamics can take much larger time steps but requires a fully converged SCF solution at each geometry, whereas each dynamical propagation step typically reaches self-consistency within a few predictor--corrector iterations, requiring only a small number of Fock-matrix constructions. Counting these constructions, which dominate the cost of both approaches, the total effort is of a similar order of magnitude for the system studied here. The main advantage demonstrated here is robustness rather than speed: the propagated trajectory follows the selected adiabatic state through a region where the static solvers either fail to converge or converge to different solutions.
However, with an optimized implementation, we expect that the dynamical propagation scheme could outperform the static SCF in terms of efficiency, especially for large condensed-phase systems.\cite{alonso2008efficient,andrade2009modified}

\subsection{Orthogonality in the DTO Representation}\label{sec:dto_results}

\begin{figure}[htbp]
    \centering
    \includegraphics[width=0.5\linewidth]{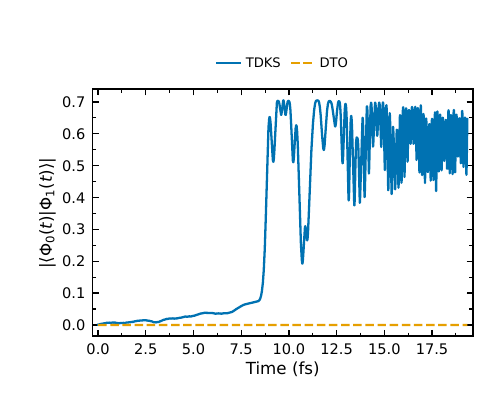}
    \caption{Magnitude of the overlap between the ground- and excited-state wavefunctions, $|\langle\Phi_{0}(t)|\Phi_{1}(t)\rangle|$, propagated with $\alpha=10$ along the RT-TDDFT/Ehrenfest path of Fig.~\ref{fig:scanning_LDA}(b). Wavefunctions constructed directly from the propagated TDKS orbitals (blue solid curve) are compared with those obtained in the DTO gauge (orange dashed curve).}
    \label{fig:DTO_overlap}
\end{figure}

Figure~\ref{fig:DTO_overlap} compares two constructions of the ground- and excited-state many-electron wavefunctions from the same TDKS orbitals propagated with $\alpha=10$ along the RT-TDDFT/Ehrenfest path of Fig.~\ref{fig:scanning_LDA}(b). When the wavefunctions are constructed directly from the propagated TDKS orbitals, their overlap stays below $0.04$ far from the crossing, rises steeply on entering the strong-coupling region, peaks at $0.70$, and fluctuates about $0.55$ thereafter. The rise coincides with the strong-coupling region identified in Sec.~\ref{sec:following}. However, in the DTO gauge, the overlap remains small: it never exceeds $1.6\times10^{-4}$ and averages $2.0\times10^{-5}$, so the two wavefunctions remain nearly orthogonal even through the strong-coupling region.

The two eigenvalues of $\mathbf{M}(t)$ remain within $10^{-4}$ of unity and zero throughout the trajectory.
These values indicate that the active space consists of one orbital of hole character, relative to the occupied ground-state space, and one of particle character, nearly orthogonal to it, although the individual propagated orbitals can mix them as the nuclei move. The DTO particle--hole construction,\cite{zhou2021dynamical} described in Sec.~\ref{sec:dto}, selects these combinations for the many-electron wavefunctions used in the overlap calculations. We repeated the comparison at $\alpha=1$, 10, 40 and 80. In each case, the DTO gauge yields similarly small overlaps while the TDKS overlap remains large. Note that near orthogonality alone is not sufficient to establish accurate adiabatic-state following: at $\alpha=1$, the DTO overlap remains small despite the substantial energy deviations seen in Sec.~\ref{sec:following}.

\subsection{Validation of the Nonadiabatic Coupling Vectors}\label{sec:nac_results}

\begin{figure}[htbp]
    \centering
        \centering
        \includegraphics[width=0.45\textwidth]{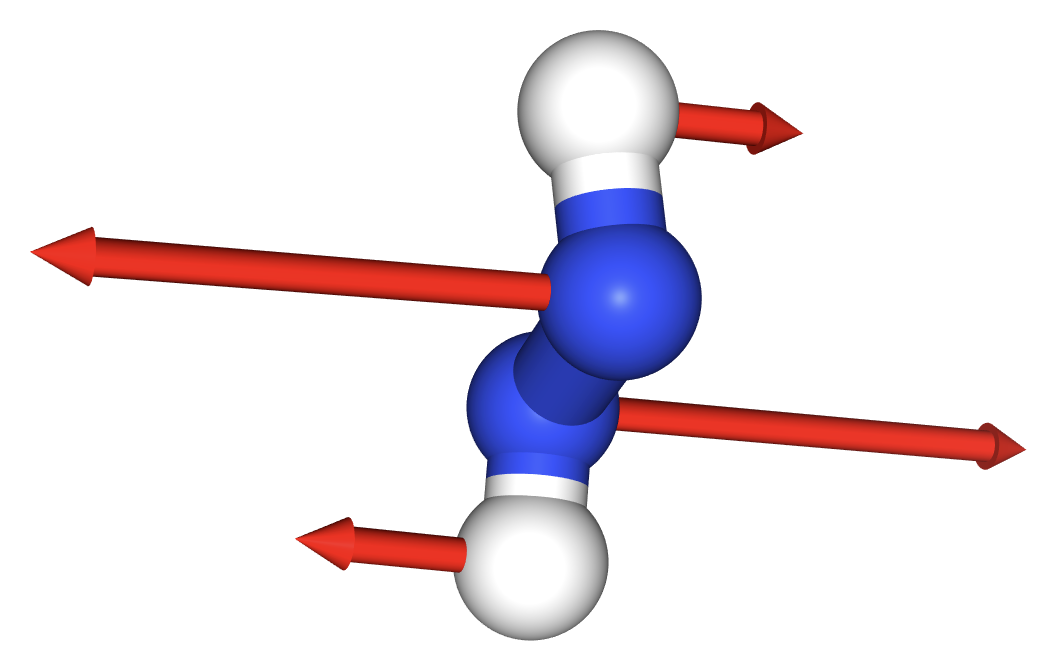}
         \caption{Spatial direction of the \textit{trans}-diimide nonadiabatic coupling vector (NACV) evaluated with the local-diabatization scheme. The Cartesian coordinates are provided in the Supporting Information.}

    \label{fig:nac_direction}
\end{figure}

We validate the finite-difference evaluation of the NACV $\boldsymbol{d}_{01}$ with RT-HF benchmark calculations on \textit{trans}-diimide. The procedure is identical for RT-TDDFT, but we chose RT-HF because its static solutions converge reliably at the displaced geometries. Both the dynamical and static vectors are evaluated by central differences with a displacement of $\Delta R=0.01$~a.u.\ ($5.3\times10^{-3}$~\AA) along each nuclear Cartesian degree of freedom, as in Eq.~\ref{eq:nacv_fd}. In both calculations, the orbital overlaps are evaluated using the AO overlap matrix between the displaced geometries, as in Eq.~\ref{eq:O_ao}. For the dynamical calculation, we fix $\alpha=1$ and vary the duration $\Delta t$ of the single Crank--Nicolson step across each displacement. The static reference is obtained from SCF solutions converged at the same displaced geometries; its direction is shown in Fig.~\ref{fig:nac_direction}. We quantify the directional agreement by the normalized dot product between the real part of the dynamically computed vector, $\boldsymbol{d}^{\text{cal}}_{01}=\operatorname{Re}(\boldsymbol{d}_{01})$, and the static reference $\boldsymbol{d}^{\text{ref}}_{01}$. Table~\ref{table:nac_convergence} shows that this dot product increases from 0.987 at $\Delta t=0.05$~a.u.\ to 0.999 at $\Delta t=4$~a.u., meaning that the direction is recovered almost perfectly.

This improvement in directional agreement with increasing $\Delta t$ is consistent with the quantum adiabatic principle of Sec.~\ref{sec:following_theory}. A single Crank--Nicolson step across a fixed displacement depends on $\alpha$ and $\Delta t$ only through their product $\alpha\Delta t$ (Eq.~\ref{eq:tau_reparam}), so increasing $\Delta t$ at $\alpha=1$ is equivalent to increasing $\alpha$ at fixed $\Delta t$. Increasing this product gives the electrons more time to respond to the displacement, bringing the propagated states closer to the adiabatic states at the displaced geometries. The propagation remains stable over the whole range examined. This comparison validates only the direction of $\boldsymbol{d}_{01}$. Its norm, $|\boldsymbol{d}_{01}|=(\boldsymbol{d}_{01}^{\dagger}\boldsymbol{d}_{01})^{1/2}$, oscillates rapidly with $\alpha$ and with the propagation interval in these tests. %
Nonetheless, it is only the direction which is required by MASH.

\begin{table}[htbp]
    \centering
    \caption{Convergence of the computed NACV direction toward the static reference with increasing propagation interval $\Delta t$, for RT-HF \textit{trans}-diimide with a finite-difference displacement $\Delta R=0.01$~a.u.\ ($5.3\times10^{-3}$~\AA) and $\alpha=1$. Alignment is quantified by the normalized dot product, in which $\boldsymbol{d}^{\text{cal}}$ is the real part $\operatorname{Re}(\boldsymbol{d}_{01})$, so that both vectors are real and $|\boldsymbol{d}|=(\boldsymbol{d}\cdot\boldsymbol{d})^{1/2}$.}
    \label{table:nac_convergence}
    \begin{tabular}{|c|c|}
        \hline
        $\Delta t$ (a.u.) & $\boldsymbol{d}^{\text{ref}}\cdot\boldsymbol{d}^{\text{cal}}/(|\boldsymbol{d}^{\text{ref}}|\,|\boldsymbol{d}^{\text{cal}}|)$ \\
        \hline
        0.05 & 0.987 \\
        0.10 & 0.988 \\
        0.40 & 0.988 \\
        0.50 & 0.988 \\
        1.00 & 0.990 \\
        2.00 & 0.995 \\
        4.00 & 0.999 \\
        \hline
    \end{tabular}
\end{table}

\subsection{Spin Dynamics on the Bloch Sphere}\label{sec:spin}

\begin{figure}[htbp]
    \centering
    \includegraphics[width=0.5\linewidth]{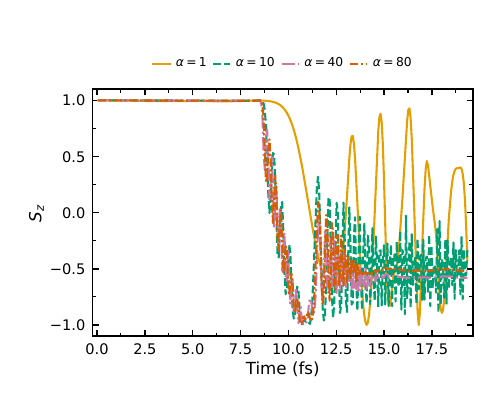}
    \caption{Evolution of the spin-mapping variable $S_z$ along the RT-TDDFT/Ehrenfest path of Fig.~\ref{fig:scanning_LDA}(b) for several values of the scaling parameter $\alpha$.} %
    \label{fig:spin_dynamics}
\end{figure}
To test whether the time-derivative couplings extracted from the propagated states drive spin-vector dynamics that converge with $\alpha$, we track $\vec{S}(t)$ on the Bloch sphere along the RT-TDDFT/Ehrenfest path of Fig.~\ref{fig:scanning_LDA}(b). The spin vector starts at the north pole, $(S_x,S_y,S_z)=(0,0,1)$, which corresponds to the first excited state ($N=1$), and the computational settings are identical to those of Fig.~\ref{fig:scanning_LDA}(b). As noted above, the nuclei follow the prescribed path, without hops or momentum rescaling. Note that the spin is propagated with the unscaled energies and couplings. Because the static CIAH ground-state solution leaves the propagated branch in the strong-coupling region (Sec.~\ref{sec:following}), overlaps between consecutive static solutions would not provide a consistent reference for the couplings there. We therefore do not include a static-reference curve in Fig.~\ref{fig:spin_dynamics}.

Figure~\ref{fig:spin_dynamics} shows $S_z$ along this path. For $\alpha\geq10$ the Bloch vector leaves the north pole near $t=8.6$~fs, as the molecule enters the strong-coupling region, and crosses the equator about 0.5~fs later. In particular, at $\alpha=1$ this transfer is delayed to 11.1~fs. However, the first-crossing times for the larger values of $\alpha$ are 9.01~fs for $\alpha=10$, 9.23~fs for $\alpha=40$ and 9.15~fs for $\alpha=80$, and are thus reasonably well converged.

After the nonadiabatic coupling region, larger values of $\alpha$ suppress the unphysical residual oscillations. With $\alpha=1$, $S_z$ keeps oscillating with an amplitude close to unity, consistent with the energy deviations seen in Sec.~\ref{sec:following}, where the propagated states depart from their adiabatic states at small $\alpha$. In particular, over the last 4~fs of the path, the standard deviation of $S_z$ falls from 0.25 at $\alpha=10$ to 0.006 at both $\alpha=40$ and 80. The averages over the same interval at $\alpha=40$ and 80 are $-0.574$ and $-0.511$, so the late-time $S_z$ is nearly converged by $\alpha=40$. Residual oscillations at small $\alpha$ could generate additional hops in fewest-switches surface hopping, where the hopping probability is computed directly from the time-derivative coupling. In MASH a hop occurs only when $S_z$ changes sign, so for hopping decisions these oscillations matter only if they carry $S_z$ across the equator.

\subsection{Nonadiabatic Dynamics}\label{sec:ensemble}

We now test the $\alpha$ dependence of the full coupled RT-TDDFT/MASH dynamics for simulating electronic relaxation and \textit{trans}--\textit{cis} isomerization. The nuclei now evolve under the force of the active electronic state instead of following the prescribed paths of Secs.~\ref{sec:following}--\ref{sec:spin}, while both electronic states are propagated at each new nuclear geometry. The active surface switches between $V_0$ and $V_1$ according to the sign of $S_z$ in Eq.~\ref{eq:mash_force_final}, and accepted hops are accompanied by momentum rescaling. Here NACV evaluations use $\alpha=1$, $\Delta R=0.01$~a.u., and a single Crank--Nicolson step of $4$~a.u.\ ($0.097$~fs) across each displacement, while the trajectory uses its own $\alpha$ and a time step of $0.4$~a.u.

The initial conditions were generated by the Wigner sampling described above, using harmonic frequencies from the ground-state LDA/6-31G* Hessian computed at its minimum-energy geometry. The resulting ensemble has an N=N bond length of $1.265\pm0.041$~\AA\ and an N--H bond length of $1.079\pm0.077$~\AA. These distributions are similar but not identical to those of the gas-phase Wigner ensemble of Mali\v{s} and Luber, which was obtained with PBE0\cite{adamo1999toward} and for which the corresponding values are $1.229\pm0.036$ and $1.037\pm0.069$~\AA.\cite{malis2020trajectory} Following the standard MASH prescription,\cite{MASH} each trajectory was initialized in the first excited state, with the initial Bloch vector sampled from the upper hemisphere of the Bloch sphere with probability density proportional to $|S_z|$.

On-the-fly RT-TDDFT/MASH simulations were then performed over a range of scaling parameters. The time-dependent excited-state population is estimated from the ensemble of time-evolved spin-mapping variables as
\begin{equation}\label{eq:mash_pop}
    P_+(t) = \frac{\langle |S_z(t=0)| \, h(S_z(t)) \rangle}{\langle |S_z(t=0)| \rangle} {}= \langle h(S_z(t))\rangle_\text{w}
\end{equation}
where the latter incorporates the weighting factor $|S_z(t=0)|$ directly into the sampling distribution. %

\begin{figure}[htbp]
    \centering
    \begin{subfigure}[t]{0.48\textwidth}
        \centering
        \includegraphics[width=\textwidth]{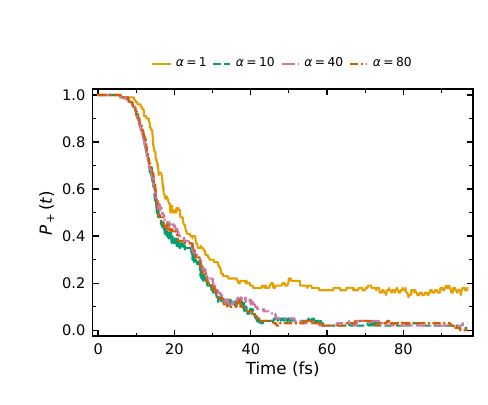}
        \caption{}
        \label{fig:mash_pop}
    \end{subfigure}
    \hfill
    \begin{subfigure}[t]{0.48\textwidth}
        \centering
        \includegraphics[width=\textwidth]{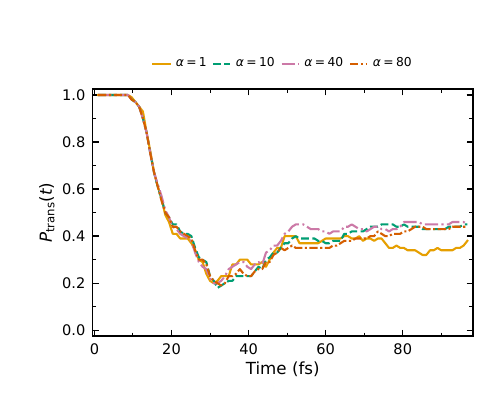}
        \caption{}
        \label{fig:ptrans}
    \end{subfigure}
    \caption{Ensemble dynamics of \textit{trans}-diimide from 100 RT-TDDFT/MASH trajectories. (a) Time-dependent excited-state population $P_{+}(t)$ of Eq.~\ref{eq:mash_pop} at several values of the scaling parameter $\alpha$. (b) Isomerization measure $P_{\mathrm{trans}}(t)$ of Eq.~\ref{eq:ptrans}.}
    \label{fig:ensemble}
\end{figure}

Figure~\ref{fig:ensemble}(a) shows that $P_+(t)$ converges once $\alpha\geq10$, with every ensemble started from the same set of initial conditions. At $\alpha=1$, the excited-state population does not decay completely, because the propagated ground state acquires considerable higher-state character, and its energy $V_0$ rises above $V_1$ for about 16\% of the time on average, compared with less than 1\% at $\alpha\geq10$. Averaging over the ensemble largely smooths out the large $S_z$ oscillations of individual $\alpha=1$ trajectories (Sec.~\ref{sec:spin}), which survive only as small fluctuations of the population about this residual value. By the end of the propagation, about $18\%$ of the ensemble is still excited at $\alpha=1$, compared with at most $3\%$ for $\alpha\geq10$.

At $\alpha=1$, the hopping histories identify frustrated relaxation and back-transfer as contributions to the residual excited-state population. Where $V_0>V_1$, transfer to the initially ground-state branch is uphill and is frustrated when the kinetic energy along the rescaling direction cannot cover the gap. The 18 trajectories remaining on the initially excited branch comprise eight with no accepted hops and ten repopulated by back-transfer. %
At $\alpha=80$, the ensemble averages about two accepted and eight frustrated hops per trajectory, with the frustrated-hop count dominated by a few trajectories with repeated attempts near $S_z=0$ (Supporting Information). Such repeated attempts are not inconsistent with converged ensemble population dynamics, as similar behavior has been reported in MASH simulations of a related system.\cite{mannouch2024quantum}

We characterize the decay by fitting the averaged population to $P_+(t)\approx P_\infty+{(1-P_\infty)e^{-(t-t_0)/\tau}}$ for $t\geq t_0$, with $P_\infty$ allowing a nonzero long-time population. The fits use nonlinear least squares, with uncertainties estimated by bootstrap resampling\cite{efron1979bootstrap} over the 100 trajectories; further details are given in the Supporting Information. The fitted parameters for all values of $\alpha$ are collected in Table~\ref{table:lifetimes}. Once the simulations are converged for $\alpha\geq10$, $t_0$ and $\tau$ agree within their statistical uncertainties. However, with $\alpha=1$, the decay starts about 2.5~fs later and proceeds faster, and the ensemble also differs significantly in its long-time limit. %

\begin{table}[htbp]
    \centering
    \caption{Fit of the averaged excited-state population to $P_+(t)\approx P_\infty+(1-P_\infty)\exp[-(t-t_0)/\tau]$ for $t\geq t_0$, for each value of $\alpha$. Uncertainties are bootstrap standard errors from 5000 resamples over the 100 trajectories of the ensemble.}
    \label{table:lifetimes}
    \begin{tabular}{|c|c|c|c|}
        \hline
        $\alpha$ & $t_0$ (fs) & $\tau$ (fs) & $P_\infty$ \\
        \hline
        1  & $11.6\pm0.7$ & $9.4\pm1.5$  & $0.169\pm0.034$ \\
        10 & $9.1\pm0.7$  & $11.4\pm1.5$ & $0.017\pm0.012$ \\
        40 & $8.8\pm0.7$  & $12.8\pm1.6$ & $0.017\pm0.014$ \\
        80 & $9.4\pm0.6$  & $11.5\pm1.4$ & $0.018\pm0.012$ \\
        \hline
    \end{tabular}
\end{table}

Mali\v{s} and Luber\cite{malis2020trajectory} quoted the lifetime as $t_0+\tau$, for which we obtain $20.9\pm1.3$~fs at $\alpha=80$. This lifetime agrees well with other simulations of the same system, although none uses exactly the same level of theory. Mali\v{s} and Luber \cite{malis2020trajectory} report a PBE0-based biexponential decay for unshifted \textit{trans}-diimide in the gas phase, with lifetimes of $37$ and $152$~fs. When they shift the $\Delta$SCF excited state down by $1.0$~eV, so that the two surfaces come almost into contact at the twisted geometry, the fitted lifetime falls to $20$~fs.\cite{malis2020trajectory} This value is relevant to our study, since the gaps in the present calculations are likewise small in the coupling region, and it is close to our result. Handt and co-workers,\cite{handt2006fragmentation} %
using an RT-TDDFT/Ehrenfest method
at the LDA/cc-pVDZ level, find the main nonadiabatic transition about 20 fs after excitation, which is also close to our result.

As in Eq.~\ref{eq:mash_pop}, we characterize isomerization by the instantaneous fraction of trajectories in the \textit{trans}-like region, defined by $|\phi_{\mathrm{HNNH}}|>90^{\circ}$,
\begin{equation}\label{eq:ptrans}
    P_{\mathrm{trans}}(t) = \frac{\langle |S_z(t=0)| \, h(|\phi_{\mathrm{HNNH}}(t)|-90^{\circ}) \rangle}{\langle |S_z(t=0)| \rangle} {}= \langle h(|\phi_{\mathrm{HNNH}}(t)|-90^{\circ})\rangle_\text{w}
\end{equation}
where $\phi_{\mathrm{HNNH}}$ is the HNNH dihedral angle. Figure~\ref{fig:ensemble}(b) shows the time-dependent \textit{trans}-like fraction $P_{\mathrm{trans}}(t)$. The four curves coincide for about the first $30$~fs, so this structural observable is less sensitive to $\alpha$ at early times than the electronic population. The subsequent departure of the $\alpha=1$ ensemble is consistent with poorer state following (Fig.~\ref{fig:scanning_LDA}). With $\alpha\geq10$, the curves agree within statistical fluctuations, and about 55\% of the ensemble occupies the \textit{cis}-like region over the last $20$~fs. Although the excited-state population has largely decayed by this time, both regions remain appreciably populated, consistent with a vibrationally excited ensemble.

\section{Conclusions} \label{sec:conclusions}

In this work, we have introduced an on-the-fly framework that combines energy-scaled \mbox{RT-TDDFT} with MASH for first-principles nonadiabatic molecular dynamics without requiring a static electronic-structure calculation at each time step. Following the quantum adiabatic principle, we scale the electronic Hamiltonian by a factor $\alpha$ to widen the separation between electronic and nuclear timescales, so that the ground- and excited-state SCF branches can be followed independently by real-time propagation. The physical energies and forces are evaluated from the unscaled functional. To use these propagated states in surface hopping, we reformulate the two-state MASH equations in a complex adiabatic basis and use a DTO gauge to construct the many-electron wavefunctions entering the overlap calculations. Local diabatization and spatial finite differences then provide the time-derivative couplings and the NACV directions required for the coupled dynamics.

We tested the method on the photoinduced dynamics of \textit{trans}-diimide. As expected, systematically increasing $\alpha$ improves agreement between the propagated energies and the static references where the latter are reliable.
Unlike the static optimizer, which produces unphysical discontinuities,
RT-TDDFT produces continuous energy surfaces through the nonadiabatic coupling region. %
The DTO construction yields nearly orthogonal wavefunctions, and the RT-HF benchmark validates the NACV direction needed for momentum rescaling. In the full MASH simulations, the electronic populations and isomerization behavior converge as $\alpha$ is increased. In particular,
increasing $\alpha$ reduces the population remaining on the initially excited branch at long times, and the lifetime is consistent with earlier simulations.\cite{malis2020trajectory,handt2006fragmentation}

This study is a proof of principle, with robustness as its main demonstrated advantage.
Extension beyond two states is possible based on the multistate formulation of MASH.\cite{unSMASH} The implementation has not been optimized for performance, and its computational efficiency relative to static SCF-based surface hopping and other nonadiabatic methods remains to be assessed. An adaptive scheme could use a large $\alpha$ only where the nonadiabaticity demands it, allowing larger time steps elsewhere. More efficient electronic propagation\cite{jia2018fast,wang2015efficient} and a multiple-time-step integrator, with the electrons advanced in substeps,\cite{tuckerman1992reversible,li2005ab,isborn2007time} would further reduce the cost. With these developments, our scheme could offer a more accurate surface-hopping alternative to RT-TDDFT/Ehrenfest simulations of large molecular and condensed-phase systems.\cite{schleife2012plane,lian2018photoexcitation,shepard2024efficient,jakowski2025simulation}

\begin{suppinfo}
Cartesian coordinates and nuclear velocities of the snapshot from which the two nuclear paths of Secs.~\ref{sec:following}--\ref{sec:spin} start; details of the fits to the excited-state population at each value of $\alpha$; convergence of the excited-state population with respect to the integration time step; hop statistics of the ensembles.
\end{suppinfo}

\section*{Data and Software Availability}

The implementation used in this work is released as an open-source Python
package, \mbox{\textsc{tdmash}}, available at
\url{https://github.com/ruiyiQM/TDMASH_pack}. The repository contains the
energy-scaled real-time propagators, the DTO gauge construction, the
nonadiabatic-coupling and MASH modules described in Secs.~\ref{sec:lcao}--\ref{sec:nac},
and input scripts that reproduce the single-trajectory and ensemble
calculations reported here. The Wigner-sampled initial conditions of the
100-trajectory ensemble are distributed with the code, so that the results of
Sec.~\ref{sec:ensemble} can be regenerated from the same starting
configurations.

\begin{acknowledgement}
R.Z. and J.O.R. thank Kasra Asnaashari and Shivank Nigam for testing some initial ideas related to this work. We further thank Yu-Chen Wang, Hannes Hoppe, Daniele Furlanetto, Jonathan R. Mannouch, Johan E. Runeson and George Trenins for helpful discussions of the MASH algorithm, and Chao Lian for advice on the implementation of RT-TDDFT.
\end{acknowledgement}

\bibliography{ref,references} %

\end{document}